\global\def\draftcontrol{0}

   \def\versionno{ dsvacuum}

\catcode`\@=11

\expandafter\ifx\csname draftcontrol\endcsname\relax\global\def\draftcontrol{0}
\fi

{\count255=\time\divide\count255 by 60
\xdef\hourmin{\number\count255}
\multiply\count255 by-60\advance\count255 by\time
\xdef\hourmin{\hourmin:\ifnum\count255<10 0\fi\the\count255}}
\def\draftdate{\number\month/\number\day/\number\year\ \ \ \hourmin }

\newcommand\makepapertitle{\par
  \begingroup
    \renewcommand\thefootnote{\@fnsymbol\c@footnote}%
    \def\@makefnmark{\rlap{\@textsuperscript{\normalfont\@thefnmark}}}%
    \long\def\@makefntext##1{\parindent 1em\noindent
            \hb@xt@1.8em{%
                \hss\@textsuperscript{\normalfont\@thefnmark}}##1}%
     \newpage
     \global\@topnum\z@   
     \@makepapertitle
     \thispagestyle{empty}\@thanks
  \endgroup
  \setcounter{footnote}{0}%
  \global\let\thanks\relax
  \global\let\makepapertitle\relax
  \global\let\@makepapertitle\relax
  \global\let\@thanks\@empty
  \global\let\@author\@empty
  \global\let\@date\@empty
  \global\let\@title\@empty
  \global\let\title\relax
  \global\let\author\relax
  \global\let\date\relax
  \global\let\and\relax
  \def\version{\let\version\@version\@gobble}
}
\def\@makepapertitle{%
  \newpage
   \ifnum\draftcontrol=1 {}
   \version\versionno
   \vskip 3em%
   \else
   \hfill\hbox to 3cm {\parbox{4cm}{\@pubnum}\hss}%
   \vskip 3em%
   \fi
   \begin{center}%
   \let \footnote \thanks
     {\LARGE {\@title}}%
     \vskip 1.5em%
     {\normalsize
       \lineskip .5em%
       \begin{tabular}[t]{c}%
         \@author
       \end{tabular}\par}%
     \vskip 1.5em%
     {\@bstract}%
     \end{center}%
     \vskip 1.5em
     \@date%
   \par
}

\gdef\@pubnum{}
\def\pubnum#1{%
  \gdef\@pubnum{#1}}

\gdef\@bstract{}
\def\Abstract#1{%
  \gdef\@bstract{%
   \parbox{\textwidth-0pc}{%
   \centerline{\bf Abstract}\penalty1000%
\kern.2cm%
\noindent
\renewcommand\baselinestretch{1.0}%
{#1}}}
}

\def\ps@paper{\let\@mkboth\@gobbletwo%
     \ifnum\draftcontrol=1
    \def\@oddfoot{\hbox to \textwidth{\tiny \versionno \hfil\tiny\draftdate}%
    \hskip -\textwidth \hbox to \textwidth{\hfil\rm\thepage\hfil}}%
     \else\def\@oddfoot{\hbox to \textwidth{\hfil\rm\thepage\hfil}}
     \fi
     \let\@evenfoot\@oddfoot
}

\def\body{\clearpage
          \pagestyle{paper}
    }

\def\@version#1{\ifnum\draftcontrol=1
\typeout{}\typeout{#1}\typeout{}
\vskip3mm\centerline{\hbox{\fbox{\normalsize{\tt DRAFT -- #1 -- }
                   {\draftdate}}}}\vskip3mm
\fi}
\let\version\@version
\long\def\eqlabel#1{\ifnum\draftcontrol=1
                    \tag@false  
                    \tag*{(\theequation) \hbox to -0.2cm{\hspace{0cm}\small{#1}\hss}}
                    \refstepcounter{equation}
                    \edef\@currentlabel{\theequation}
                    \ltx@label{#1}          
                    \else
                    \label{#1}
                    \fi
                    }
\let\st@bibitem\@bibitem
\let\st@lbibitem\@lbibitem
\ifnum\draftcontrol=1
  \def\@bibitem#1{%
    \st@bibitem{#1}\a@@label{#1}\ignorespaces}
  \def\@lbibitem[#1]#2{%
    \st@lbibitem[#1]{#2}\a@@label{#2}\ignorespaces}
  \def\a@@label#1{%
    \gdef\a@lab{\smash{\normalfont\small#1}}
    \ifvmode
      \if@inlabel
        \global\setbox\@labels\hbox{%
          \llap{\a@lab\let\a@lab\relax
                \kern\@totalleftmargin\kern\marginparsep}%
          \box\@labels}%
      \fi
    \fi}
\fi

\documentclass[12pt,letterpaper]{article}

\usepackage{amsmath,amssymb,array,calc,epsfig,rotating,bm}
\usepackage[sort]{cite}
\usepackage{graphicx}
\usepackage{psfrag,verbatim}

\ifnum\draftcontrol=1
\fi

\renewcommand\baselinestretch{1.25}
\renewcommand\section{\@startsection {section}{1}{\z@}%
                                   {-3.5ex \@plus -1ex \@minus -.2ex}%
                                   {2.3ex \@plus.2ex}%
                                   {\normalfont\large\bfseries}}
\renewcommand\subsection{\@startsection{subsection}{2}{\z@}%
                                   {-3.25ex\@plus -1ex \@minus -.2ex}%
                                   {1.5ex \@plus .2ex}%
                                   {\normalfont\normalsize\bfseries}}
\renewcommand\subsubsection{\@startsection{subsubsection}{3}{\z@}%
                                   {-3.25ex\@plus -1ex \@minus -.2ex}%
                                   {1.5ex \@plus .2ex}%
                                   {\normalfont\normalsize\it}}
\renewcommand\paragraph{\@startsection{paragraph}{4}{\z@}%
                                   {-3.25ex\@plus -1ex \@minus -.2ex}%
                                   {1.5ex \@plus .2ex}%
                                   {\normalfont\normalsize\bf}}

\numberwithin{equation}{section}

\def\revise#1       {\raisebox{-0em}{\rule{3pt}{1em}}%
                     \marginpar{\raisebox{.5em}{\vrule width3pt\
                     \vrule width0pt height 0pt depth0.5em
                     \hbox to 0cm{\hspace{0cm}{%
                     \parbox[t]{4em}{\raggedright\footnotesize{#1}}}\hss}}}}

\newcommand\nxt[1]  {\\\fnxt#1}
\newcommand{\ie}{{\it i.e.,}\ }

\def\cala         {{\cal A}}

\def\cale         {{\cal E}}

\def\call         {{\cal L}}
\def\calm         {{\cal M}}
\def\caln         {{\cal N}}
\def\calo         {{\cal O}}
\def\calp         {{\cal P}}

\def\calr         {{\cal R}}

\def\reals        {{\mathbb R}}

\def\del          {\partial}

\def\tr           {\mathop{\rm Tr}}

\def\sqr#1#2{{\vcenter{\vbox{\hrule height.#2pt
 \hbox{\vrule width.#2pt height#1pt \kern#1pt
 \vrule width.#2pt}\hrule height.#2pt}}}}

\def\a{\alpha}
\def\b{\beta}

\def\r{\rho}
\def\dd{\delta}

\def\c{\chi}

\def\aa1{\phi}
\def\cc1{\psi}

\def\l{\lambda}

\def\t{\tau}

\def\tQ{\tilde{Q}}

\catcode`\@=12

\begin{document}


\title{\bf de Sitter Vacua of Strongly Interacting QFT}

\date{February 4, 2017}

\author{
Alex Buchel$ ^{1,2,3}$ and Aleksandr Karapetyan$ ^1$\\[0.4cm]
\it $ ^1$Department of Applied Mathematics\\
\it $ ^2$Department of Physics and Astronomy\\ 
\it University of Western Ontario\\
\it London, Ontario N6A 5B7, Canada\\
\it $ ^3$Perimeter Institute for Theoretical Physics\\
\it Waterloo, Ontario N2J 2W9, Canada
}

\Abstract{We use holographic correspondence to argue that Euclidean
(Bunch-Davies) vacuum is a late-time attractor of the dynamical
evolution of quantum gauge theories at strong coupling.  The
Bunch-Davies vacuum is not an adiabatic state, if the gauge theory is
non-conformal --- the comoving entropy production rate is nonzero.
Using the $\caln=2^*$ gauge theory holography, we explore prospects of
explaining current accelerated expansion of the Universe as due to the
vacuum energy of a strongly coupled QFT.
}

\makepapertitle

\body

\version\versionno
\tableofcontents

\section{Introduction}\label{intro}
One of the outstanding questions of cosmology is the physics behind
the current accelerated expansion of the Universe
\cite{Riess:1998cb,Perlmutter:1998np}.
We argue that
there is a strong coupling aspect of this question, which can be addressed
using the holographic correspondence between strongly coupled
gauge theories and dual gravitational models in asymptotically
anti-de-Sitter (AdS) space-times \cite{m1,Aharony:1999ti}. 

In a framework
of classical Einstein gravity and quantum field theory,
the dynamics of the Universe is described
by the following equation
\begin{equation}
G_{\mu\nu} = 8\pi G_N\ \langle T_{\mu\nu}\rangle\,,
\eqlabel{einstein}
\end{equation}
where $G_{\mu\nu}$ is the Einstein tensor of the four-dimensional space-time
metric $g_{\a\b}^{(4)}$, $G_N=M_{pl}^{-2}$ is  Newton's constant, and
$\langle T_{\mu\nu}\rangle =\langle
T_{\mu\nu}[g_{\a\b}^{(4)}]\rangle $ is the quantum expectation value
of the stress-energy tensor 
of all matter fields, self-consistently computed in the space-time
background metric $g_{\a\b}^{(4)}$. Gravitational Bianchi identity
implies covariant conservation of the stress-energy tensor:
\begin{equation}
\nabla^{\mu} \langle T_{\mu\nu}\rangle =0\,.
\eqlabel{constmunu}
\end{equation}
At late times, $t\to \infty$,
expansion of the Universe
will ultimately dilute all the components of the stress energy tensor,
except for the vacuum energy,
\begin{equation}
\lim_{t\to \infty} \langle T_{\mu\nu}\rangle = \langle {\rm vacuum}
| T_{\mu\nu}|{\rm vacuum}\rangle ={\rm diag }(\cale_v,P_v,P_v,P_v)\,,
\eqlabel{latetimes}
\end{equation}
where $\cale_v$ and $P_v$ are the matter fields vacuum energy density and pressure correspondingly,
where, using \eqref{constmunu},
\begin{equation}
\cale_v+P_v=0\,.
\eqlabel{epcons}
\end{equation}
If $\cale_v>0$, the (spatially-flat) Universe at late-times is de Sitter, with a Hubble constant $H$:
\begin{equation}
ds_4^2=g_{\a\b}^{(4)} dx^\a dx^\b = -dt^2 +a(t)^2\ d\boldsymbol{x}^2\,,\qquad a(t)=e^{Ht}\,,
\eqlabel{desitter1}
\end{equation}
such that
\begin{equation}
H^2 M_{pl}^2=\frac{8\pi}{3}\ \cale_v(H)\,.
\eqlabel{desitter2}
\end{equation}
where we explicitly indicated that the vacuum energy of the Standard Model (SM) of particle physics 
should be computed self-consistently in \eqref{desitter1}. Values of $H$ obtained from 
solving \eqref{desitter2} can be reliable in so far as approximation \eqref{einstein} is valid, \ie
$H\ll M_{pl}$, so that the effects of quantum gravity can be neglected. 

Computation of a vacuum energy of a QFT in curved space-time is a subject with a long history
\cite{Birrell:1982ix}. Almost all analysis are done though in the framework of perturbative 
{\it weakly-coupled} QFT. Irrespectively of the physics beyond the SM, the SM itself contains 
a sector with a {\it strongly coupled} sector in the infrared --- namely, quantum chromodynamics (QCD).  
This is the main motivating factor of the present project --- we use the holographic 
correspondence to compute the vacuum energy of certain strongly coupled gauge theories in de Sitter 
space-time. 

Holographic computations of de Sitter vacuum energy of conformal field theories 
at strong coupling was done previously in \cite{Hawking:2000bb}. However, because the 
theory in question was conformal, $\cale_v\sim H^4$, since the Hubble constant is the only scale parameter 
in the model. Necessarily,  \eqref{desitter2} implies $H\sim M_{pl}$ which invalidates 
classical treatment of gravity in this regime. To have a chance of a self-consistent treatment 
of a classical Einstein gravity coupled to quantum SM, the latter should be modeled 
as a non-conformal gauge theory in a holographic setting. For one reason, QCD is responsible 
for a strongly coupled vacuum in our Universe precisely because it is non-conformal. 
Second, existence of an intrinsic scale $\Lambda$ in quantum gauge theory allows for a 
potential hierarchy between the scales $H$ and $M_{pl}$ generated in a relation \eqref{desitter2}.

Non-conformal gauge theories in de Sitter background were studied in holographic framework 
earlier \cite{Buchel:2002wf,Buchel:2002kj,Buchel:2003qm,Buchel:2004qg,Buchel:2006em,Buchel:2013dla,Anguelova:2014dza,Anguelova:2015dgt}. 
The basic idea was to take a holographic correspondence describing a renormalization group (RG) flow for a vacuum state of 
a strongly coupled non-conformal gauge theory in Minkowski space-time and ``deform'' it to a background de Sitter geometry,
 $\reals^{3,1}\to dS_4$. The resulting de Sitter RG flow, identified as a vacuum state of the corresponding 
gauge theory in $dS_4$,  exhibited the $SO(4,1)$ symmetry and allowed for a Euclidean 
analytical continuation.  Thus, all the vacua discussed in \cite{Buchel:2002wf,Buchel:2002kj,Buchel:2003qm,Buchel:2004qg,Buchel:2006em,Buchel:2013dla}
are Euclidean or Bunch-Davies \cite{Bunch:1978yq}. It was recognized in \cite{Mottola:1984ar,Allen:1985ux} (MA)
that the vacuum state invariance under the proper de Sitter transformations in free field theories allows for a 
two-parameter family of inequivalent vacua. These MA-vacua are interesting in that they might lead to enhanced, 
order $\propto \frac{H}{M_{pl}}$, effects during inflation \cite{Easther:2001fi}. Subsequent studies of 
perturbative interactions in MA-vacua revealed various problems \cite{Banks:2002nv,Einhorn:2002nu,Einhorn:2003xb,Collins:2003zv}.
To our knowledge, no corresponding studies at strong coupling were ever performed. As a second motivation 
for our project, we use holographic framework to study the evolution of a generic spatially homogeneous and isotropic 
state in strongly coupled non-conformal gauge theory. We find that while at any finite boundary time $t$ 
the state is not invariant under $dS_4$ isometries, in the infinite future $t\to \infty$ the state evolves to a 
Bunch-Davies vacuum, equivalent to the ones discussed in  \cite{Buchel:2002wf,Buchel:2002kj,Buchel:2003qm,Buchel:2004qg,Buchel:2006em,Buchel:2013dla}. 
A corollary of our analysis is that MA-vacua can not be reached dynamically from homogeneous and isotropic initial 
conditions in strongly coupled gauge theories with a holographic dual. This further suggests that MA-vacua are inconsistent 
in interacting quantum gauge theories.      

There are several ways to generate  non-conformal RG flows in holography. 
One can start with $\caln=4$ superconformal Yang-Mills theory $AdS_5\times S^5$ duality \cite{m1} 
and deform it by turning on a nonzero coupling for a relevant operator, explicitly breaking the scale invariance
of the parent theory. A typical representative here is the $\caln=2^*$ holography \cite{Pilch:2000ue,Buchel:2000cn,Evans:2000ct}.
Alternatively, the scale invariance of the parent theory can be broken spontaneously, as it is in the case 
of Klebanov-Strassler (KS) cascading gauge theory \cite{Klebanov:2000hb}. Finally, holography for a  
non-conformal gauge theory in four space-times dimensions can be obtained from the 
Kaluza-Klein reduction of the higher-dimensional scale invariant gauge theory/gravity correspondence, 
as in    \cite{Witten:1998zw}.
In this paper we focus on the $\caln=2^*$ holographic correspondence and delegate the discussion of the KS theory
to future work\footnote{Unlike non-conformal gauge theories obtained from compactifications 
of the higher-dimensional conformal gauge theories, KS cascading 
gauge theory is intrinsically four-dimensional \cite{Aharony:2005zr}. }.    

The rest of the paper is organized as follows. 
In the next section we briefly review the holographic duality between 
$\caln=2^*$ supersymmetric $SU(N)$ gauge theory \cite{Buchel:2000cn} 
and its Pilch-Warner (PW) gravitational dual  \cite{Pilch:2000ue}.
In section \ref{evolve},  building on \cite{Buchel:2016cbj}, we discuss evolution of generic 
homogeneous and isotropic states of $\caln=2^*$ gauge theory in $dS_4$. 
 In section \ref{venergy} we show that at late times these states approach a Bunch-Davies (Euclidean) vacuum of the theory 
constructed in \cite{Buchel:2003qm}. 
We argue that the Euclidean vacuum of non-conformal gauge theories is not an adiabatic state --- the comoving entropy 
continues to be produced.
We compute the vacuum energy 
$\cale_v=\cale_v(H)$ and the entropy production rate of strongly coupled $\caln=2^*$ gauge theory in $dS_4$.   
We conclude in section \ref{conclude} with speculations regarding explanation of 
the current accelerated expansion of the Universe due to a vacuum energy 
of a gauge theory sector at strong coupling\footnote{This proposal originated 
from discussions with David Mateos and Jorge Casalderrey-Solana.}.

\section{$\caln=2^*$ holography}\label{n2holography}
 
$\caln=2^*$ holography is a correspondence between mass-deformed $\caln=4$ $SU(N)$ supersymmetric 
Yang-Mill theory at strong coupling and a PW five-dimensional $\caln=2$ supergravity --- a particular consistent truncation of 
ten-dimensional type IIb supergravity. 

\subsection{$\caln=2^*$ gauge theory}

In the language of four-dimensional $\caln=1$ supersymmetry, the
mass deformed $\caln=4$ $SU(N)$ Yang-Mills theory ($\caln=2^*$) in
$\reals^{3,1}$ consists of a vector multiplet $V$, an adjoint chiral
superfield $\Phi$ related by $\caln=2$ supersymmetry to the gauge
field, and two additional adjoint chiral multiplets $Q$ and $\tilde{Q}$
which form an $\caln=2$ hypermultiplet.  In addition to the usual
gauge-invariant kinetic terms for these fields%
\footnote{The classical K\"{a}hler potential is normalized
according to $(2/g_{YM}^2)\tr[\bar{\Phi}\Phi+ \bar{Q}Q+\bar{\tQ}\tQ]$.},
the theory has additional interactions and a hypermultiplet mass term
given by the superpotential
\begin{equation}
W=\frac{2\sqrt{2}}{g_{YM}^2}\tr([Q,\tQ]\Phi)
+\frac{m} {g_{YM}^2}(\tr Q^2+\tr\tQ^2)\,.
\eqlabel{sp}
\end{equation}
When $m=0$, the gauge theory is superconformal with $g_{YM}$
characterizing an exactly marginal deformation. The theory has a
classical $3(N-1)$ complex dimensional moduli space, which is
protected by supersymmetry against (non)-perturbative quantum
corrections.

When $m\ne 0$, the $\caln=4$ supersymmetry is softly broken to
$\caln=2$. This mass deformation lifts the $\{Q,\ \tQ\}$ hypermultiplet
moduli directions, leaving the $(N-1)$ complex dimensional Coulomb
branch of the $\caln=2$ $SU(N)$ Yang-Mills theory, parameterized by
expectation values of the adjoint scalar
\begin{equation}
\Phi={\rm diag} (a_1,a_2,\cdots,a_N)\,,\quad \sum_i a_i=0\,,
\eqlabel{adsc}
\end{equation}
in the Cartan subalgebra of the gauge group.  For generic values
of the moduli $a_i$, the gauge symmetry is broken to that of the
Cartan subalgebra $U(1)^{N-1}$, up to the permutation of individual
$U(1)$ factors. Additionally, the superpotential \eqref{sp} induces
the RG flow of the gauge coupling.  While from the gauge theory
perspective it is straightforward to study this $\caln=2^{*}$ theory
at any point on the Coulomb branch \cite{Donagi:1995cf}, the PW supergravity
flow \cite{Pilch:2000ue} corresponds to a particular Coulomb branch vacuum.
More specifically, matching the probe computation in gauge theory
and the dual PW supergravity flow, it was argued in \cite{Buchel:2000cn} that
the appropriate Coulomb branch vacuum corresponds to a  linear
distribution of the vevs \eqref{adsc} as
\begin{equation}
a_i\in [-a_0,a_0]\,,\qquad a_0^2=\frac{m^2 g_{YM}^2 N}{\pi}\,,
\eqlabel{inter}
\end{equation}
with (continuous in the large $N$ limit) linear number density
\begin{equation}
\rho(a)=\frac{2}{m^2 g_{YM}^2}\sqrt{a_0^2-a^2}\,,\qquad
\int_{-a_0}^{a_0}da \rho(a)=N\,.
\eqlabel{rho}
\end{equation}
The special character of the Coulomb branch vacuum \eqref{rho} was explained in  \cite{Buchel:2013id},
where it was shown that this vacuum is a saddle point of the supersymmetric localization of 
$\caln=2^*$ gauge theory on $S^4$ ( in the $S^4$  decompactification limit ) in the planar limit, 
\ie $N\to \infty$ and  $g_{YM}^2\to 0$ with the 't Hooft coupling $N g_{YM}^2={\rm const}\gg 1$.  

\subsection{PW effective action}

The gravitational dual to $\caln=2^*$ supersymmetric gauge theory in the planar limit 
and for large 't Hooft coupling $N g_{YM}^2\gg 1$ was constructed in \cite{Pilch:2000ue}. 
A consistent truncation from type IIb supergravity is encoded in five-dimensional 
effective action 
\begin{equation}
\begin{split}
S=&\,
\int_{\calm_5} d\xi^5 \sqrt{-g}\ \call_5\\
=&\frac{1}{16\pi G_5}\,
\int_{\calm_5} d\xi^5 \sqrt{-g}\left[R-12 (\del\a)^2-4 (\del\chi)^2-
\calp\right]\,,
\end{split}
\eqlabel{action5}
\end{equation}
where the potential%
\footnote{We set the five-dimensional gauged
supergravity coupling to one. This corresponds to setting the
radius $L$ of the five-dimensional sphere in the undeformed metric
to $2$.}
\begin{equation}
\calp=\frac{1}{4}\left[\frac 13 \left(\frac{\del W}{\del
\a}\right)^2+ \left(\frac{\del W}{\del \chi}\right)^2\right]-\frac
43 W^2\,
 \eqlabel{pp}
\end{equation}
is a functional of supergravity scalars $\alpha$ and $\chi$, and is determined by the
superpotential
\begin{equation}
W=- e^{-2\alpha} - \frac{1}{2} e^{4\alpha} \cosh(2\chi)\,.
\eqlabel{supp}
\end{equation}
In our conventions, the five-dimensional Newton's constant is
\begin{equation}
G_5\equiv \frac{G_{10}}{2^5\ {\rm vol}_{S^5}}=\frac{4\pi}{N^2}\,.
\eqlabel{g5}
\end{equation}
Effective action \eqref{action5} can be further consistently truncated to 
$\{\chi=0\,,\ \a\ne 0\}$ or to $\{\a=\c=0\}$. In the later case the vacuum solution is that of 
$AdS_5$, holographically dual to a vacuum of $\caln=4$ SYM. 

The gravitational scalars $\a$ and $\chi$ are dual to dimension-two $\calo_2$ 
and dimension-three $\calo_3$ operators of $\caln=4$ SYM,
\begin{equation}
\begin{split}
&\a\Longleftrightarrow\ \calo_2=\frac13 {\tr}\left(\, |\phi_1|^2 + |\phi_2|^2 - 2\,|\phi_3|^2
\,\right)\,,
\\
&\c\Longleftrightarrow\ \calo_3=-{\tr}\left( i\,\psi_1\psi_2 -\sqrt{2}g_{YM}\,\phi_3
[\phi_1,\phi_1^\dagger] +\sqrt{2}g_{YM}\,\phi_3
[\phi_2^\dagger,\phi_2] + {\rm h.c.}\right)\\
&\qquad\qquad +\frac23 m_f\, {\tr}\left(\, |\phi_1|^2 + |\phi_2|^2 +
|\phi_3|^2\, \right)\,,
\end{split}
\eqlabel{o2o3}
\end{equation} 
where $\phi_{1,2}$ are the bosonic components of the $\caln=2$ hypermultiplet 
chiral superfields $\{Q,\tQ\}$, and $\phi_3$ is a bosonic component of the chiral 
superfield $\Phi$; $\psi_i$ are the fermionic superpartners of $\phi_i$. 
The fermionic mass term $m_f$ and the bosonic mass term $m_b^2$
(the nonnormalizable coefficients of $\c$ and $\a$ near the $AdS_5$ boundary correspondingly)
realize the $\caln=4$ SYM theory deformation
\begin{equation}
\delta \call= -2\,\int d^4x\,\left[ \,m_b^2\,\calo_2\,
+m_f\,\calo_3\,\right]\,. 
\eqlabel{massdef}
\end{equation}
For a generic choice $m_f^2\ne m_b^2$ this deformation completely breaks 
the supersymmetry; when $m_f^2=m_b^2\equiv m^2$, $\dd\call$ is the mass deformation 
in the $\caln=2$ supersymmetric superpotential \eqref{sp}.   

PW effective action \eqref{action5} does not reproduce the free energy of $\caln=2^*$ gauge theory 
on $S^4$ computed from the supersymmetric localization  in the $S^4$ decompactification 
limit \cite{Buchel:2013fpa}. Rather, additional couplings to the background $S^4$ metric are needed
\cite{Pestun:2007rz}, resulting in an enlarged gravitational dual \cite{Bobev:2013cja} (BEFP).  
PW effective action is a consistent truncation of the BEFP effective 
action. 
As outlined in the introduction, we are interested in the dynamics of $\caln=2^*$ gauge theory 
in $dS_4$. Since the gauge theory background 
space-time of interest is curved, the BEFP effective action encoding additional 
curvature couplings, allows for a more general 
analysis. For simplicity, in this paper we limit the scope of 
study to PW truncation\footnote{It would be interesting 
to explore the parameter space of the curvature couplings, 
and study the stability of the PW 
truncation within BEFP for $\caln=2^*$ gauge theory formulated in curved space-time, 
extending stability analysis for thermal states of $\caln=2^*$ plasma in Minkowski space-time 
\cite{Balasubramanian:2013esa}.}.

\section{Holographic evolution of $\caln=2^*$ gauge theory states in $dS_4$}\label{evolve}

We use the characteristic formulation of gravitational dynamics in
asymptotically AdS space-times summarized in \cite{Chesler:2013lia} to
describe evolution of spatially homogeneous and isotropic  states of $\caln=2^*$ gauge theory in
$dS_4$. We follow closely the discussion in \cite{Buchel:2016cbj}.

A generic state of the gauge theory, homogeneous and isotropic in the spatial
boundary coordinates $\boldsymbol{x}=\{x,y,z\}$, leads to a dual gravitational metric ansatz
\begin{equation}
ds_5^2=2 dt\ (dr -A dt) +\Sigma^2\ d\boldsymbol{x}^2\,,
\eqlabel{EFmetric}
\end{equation}
with the warp factors $A,\Sigma$ as well as the bulk scalars $\a,\chi$
depending only on $\{t,r\}$. From
PW effective action \eqref{action5} we obtain the following equations of
motion:
\begin{equation}
\begin{split}
&0=d_+'\Sigma+2 {\Sigma'}\ d_+\ln\Sigma+
\frac \Sigma6\ \calp,\\ 
&0=A''-6(\ln\Sigma)'\ d_+\ln\Sigma +4\chi' d_+\chi +12\a' d_+\a 
-\frac \calp6, \\ 
&d_+'\a+\frac{3}{2}\ \left((\ln\Sigma)'d_+\a+\a' d_+\ln\Sigma \right) 
-\frac {1}{48} \del_\a \calp, \\ 
&0=d_+'\chi+\frac{3}{2}\ 
\left((\ln\Sigma)'d_+\chi+\chi' d_+\ln\Sigma \right)-\frac{1}{16}\del_\chi \calp,
\end{split}
\eqlabel{ev1}
\end{equation}
as well as the Hamiltonian constraint equation:
\begin{equation}
0=\Sigma''+\left(4 (\a')^2+\frac 43 (\chi')^2\right) \Sigma\,,
\eqlabel{ham}
\end{equation}
and the momentum constraint equation:
\begin{equation}
\begin{split}
&0=d^2_+\Sigma -2 A\Sigma' -(4 A \Sigma'+A' \Sigma)d_+\ln\Sigma
\\
&+\left(4 (d_+\a)^2+\frac 43 (d_+\chi)^2\right)\Sigma  
-\frac 13 \Sigma A \calp\,.
\end{split}
\eqlabel{mom}
\end{equation}
In \eqref{ev1}-\eqref{mom} 
we denoted $'= \frac{\del}{\del r}$, $\dot\ =\frac{\del}{\del t}$, 
and $d_+= \frac{\del}{\del t}+A \frac{\del }{\del r}$. 
The near-boundary $r\to\infty$ asymptotic behaviour\footnote{As explained in \cite{Chesler:2013lia},
the asymptotic expansion is determined up to a radial coordinate shifts $r\to r+\l(t)$,
with an arbitrary function $\l(t)$. We fixed this residual reparametrization specifying the
the $\calo(r)$ term in the asymptotic expansion for the metric function $A$.
We verified that the physical observables are independent of the reparametrization choice.}
of the metric
function and the scalars encode the mass parameters $m_b^2$ and $m_f$ of
the $\caln=2^*$ gauge theory \cite{Buchel:2007vy}, and the boundary
metric \eqref{desitter1} scale factor $a(t)$:
\begin{equation}
\begin{split}
&\Sigma=\frac{ar}{2}+\calo(r^{-1})\,,\qquad A=\frac{r^2}{8}-\frac{\dot a r}{a }+\calo(r^0)\,,\\
&\a=-\frac{8m_b^2\ln r}{3 r^2}+\calo(r^{-2})\,,\qquad \chi=\frac{2m_f}{r}+\calo(r^{-2})\,.
\end{split}
\eqlabel{bcdata}
\end{equation}
An initial state of the gauge theory is specified providing the scalar
profiles $\a(0,r)$ and $\chi(0,r)$ and solving the
constraint \eqref{ham}, subject to the boundary
conditions \eqref{bcdata}. Equations \eqref{ev1} can then be used to evolve
the state. Note that the gravitational bulk metric \eqref{EFmetric}
does not enjoy the $SO(4,1)$ boundary metric isometry.

The subleading terms in the boundary expansion of the
metric functions and the scalars encode the evolution of the  energy
density $\cale(t)$, pressure $P(t)$ and the expectation values of the operators
$\calo_2(t)$ and $\calo_3(t)$ of the prescribed gauge theory initial state.
Specifically, extending to asymptotic expansion \eqref{bcdata} for $\{\a,\c, A\}$,
\begin{equation}
\begin{split}
&\a=-\frac{8m_b^2\ln r}{3 r^2}+\frac{\a_{2,0}(t)}{r^2}+\calo\left(\frac{\ln r}{r^3}\right)\,,\\
&\c=\frac{2 m_f}{r}+\frac{8 \dot{a} m_f}{a r^2}+\frac{1}{r^3}\biggl(\c_{3,0}(t)-\left(
\frac{32}{3} m_f^3+16 m_f \frac{\ddot{a}}{a}+16 m_f \frac{(\dot{a})^2}{a^2}\right) \ln r
\biggr)+\calo\left(\frac{\ln r}{r^4}\right)\,,\\
&A=\frac{r^2}{8}-\frac{\dot{a} r}{a}-\frac23 m_f^2
+\frac{1}{r^2}\biggl(
a_{2,0}(t)+\biggl(
\frac{32}{9} m_f^4+\frac{16}{9} m_b^2 \a_{2,0}(t)-\frac{32}{81} m_b^4
+\frac{32\ddot{a}}{3a} m_f^2
\biggr) \ln r\\
&-\frac{64}{27} m_b^4 \ln^2 r
\biggr)+\calo\left(\frac{\ln^2 r}{r^3}\right)\,,
\end{split}
\eqlabel{extension}
\end{equation}
the observables of interest can be computed following the
holographic renormalization of the model, most recently reviewed  in
\cite{Buchel:2012gw},
\begin{equation}\begin{split}
\frac{32\pi^2}{N^2} \cale(t)=&-\frac38 a_{2,0}+3 \frac{(\dot a)^4}{a^4}
-\frac14 \chi_{3,0} m_f-\frac 18 \a_{2,0}^2
+\frac59 m_b^2 \a_{2,0} 
-\left(\frac 83 \ln 2+\frac{40}{81}\right) m_b^4
\\&+\left( \frac 83 \ln 2+\frac 79\right) m_f^4
+m_f^2 \biggl(
\frac{16\ddot a}{3a} +\frac{(\dot a)^2}{a^2} (8 \ln 2+6)\biggr)
+\calo_\cale^{ambiguity}\,,
\end{split}
\eqlabel{vev1}
\end{equation}
\begin{equation}\begin{split}
\frac{32\pi^2}{N^2} P(t)=&-\frac18 a_{2,0}
-\frac{4 (\dot a)^2 \ddot a}{a^3}
+\frac{(\dot a)^4}{a^4}+\frac{1}{12} \chi_{3,0} m_f
-\frac{1}{24} \a_{2,0}^2
-\frac{13}{27} m_b^2 \a_{2,0}
\\&+\left(\frac 83 \ln 2+\frac{68}{243}\right) m_b^4
+\left(\frac{7}{27}-\frac83 \ln 2\right) m_f^4
-m_f^2 \biggl(
\left(\frac{16}{3} \ln 2+\frac89\right) \frac{\ddot a}{a}
\\
&+\left(\frac83 \ln 2+2\right) \frac{(\dot a)^2}{a^2}
\biggr)
+\calo_P^{ambiguity}\,,
\end{split}
\eqlabel{vev2}
\end{equation}
\begin{equation}
\begin{split}
\frac{32\pi^2}{N^2} \calo_3(t)=&-\frac14 \chi_{3,0}
+\left(\frac{16}{3} \ln 2-\frac 23\right) m_f^3
+m_f \left((8 \ln 2+2) \frac{\ddot a}{a}+(8 \ln 2+4) 
\frac{(\dot a)^2}{a^2}
\right)\\
&+\calo_3^{ambiguity}\,,
\end{split}
\eqlabel{vev3}
\end{equation}
\begin{equation}\begin{split}
\frac{32\pi^2}{N^2} \calo_2(t)=&2 \a_{2,0}-\frac{32}{3} m_b^2 \ln 2
+\calo_2^{ambiguity}\,,
\end{split}
\eqlabel{vev4}
\end{equation}
where $\calo^{ambiguity}$ denote renormalization-scheme dependent  
ambiguities of the expectation values, parameterized by $\dd_1,\cdots \dd_4$, \cite{Buchel:2012gw}:
\begin{equation}
\begin{split}
\calo^{ambiguity}_\cale=&\frac 14 m_f^4 \delta_1+\frac19 m_b^4 \delta_4
+\frac{(\dot a)^2}{a^2} \left(\delta_2 m_b^2+\frac 32 \delta_3 m_f^2\right)\,,
\cr
\calo^{ambiguity}_P=&-\frac14 m_f^4 \delta_1
-\frac19 m_b^4 \delta_4
-\left(\frac{2 \ddot a}{a}+\frac{(\dot a)^2}{a^2}\right)
\left(\frac 13 \delta_2 m_b^2+\frac 12 \delta_3 m_f^2\right)\,,
\\
\calo^{ambiguity}_3=&\frac12 \delta_1 m_f^3
+\frac32 m_f \delta_3 \left( \frac{\ddot a}{a}+\frac{(\dot a)^2}{a^2}
\right)\,,
\cr
\calo^{ambiguity}_2=&\frac 49 \delta_4 m_b^2
+2 \delta_2 \left( \frac{\ddot a}{a}+\frac{(\dot a)^2}{a^2}\right)\,.
\end{split}
\eqlabel{vevsamb}
\end{equation}
Independent of the regularization scheme parameters $\dd_1,\cdots
, \dd_4$, these expectation values 
satisfy the expected conformal anomaly relation
\begin{equation}
\begin{split}
&-\cale+3P=\frac{N^2}{32\pi^2}\left(R_{\mu\nu}R^{\mu\nu}-\frac 13 R^2\right)-2 m_f \calo_3-m_b^2 \calo_2 
-\frac{N^2}{24\pi^2}\left(m_f^4-m_b^4+\frac {m_f^2}{2} R\right)\,,
\end{split}
\eqlabel{ward}
\end{equation}
Furthermore, the conservation of the stress-energy tensor 
\begin{equation}\eqlabel{cons}
\frac{d\cale}{dt}+3\frac{\dot a}{a} (\cale+P)=0\,,
\end{equation}
is a consequence of the momentum constraint \eqref{mom}:   
\begin{equation}
\begin{split}
&0=\dot a_{2,0}+\frac{4 \dot a a_{2,0}}{a}
+\frac23 \a_{2,0} \dot \a_{2,0}+\frac{4\dot a \a_{2,0}^2}{3a}
+\left( \frac{416}{243} m_b^4-\frac{224}{27} m_f^4 \right) 
\frac{\dot a}{a}
\\&-\left(
\frac{160\dot a \ddot a}{3a^2}
+\frac{128\dddot a}{9a}
\right) m_f^2
-\left(
\frac{16\dot a \a_{2,0}}{27a}+\frac{40\a_{2,0}}{27}
\right) m_b^2
+\left(
\frac{4\dot a \chi_{3,0}}{3a}+\frac{2\dot\chi_{3,0} }{3} 
\right) m_f\,.
\end{split}
\eqlabel{ward2}
\end{equation}

One of the advantages of the holographic formulation of a QFT dynamics is the natural definition 
of its far-from-equilibrium entropy density. A gravitational geometry \eqref{EFmetric} 
has an apparent horizon located at $r=r_{AH}$, where \cite{Chesler:2013lia}
\begin{equation}
d_+\Sigma\bigg|_{r=r_{AH}}=0\,.
\eqlabel{defhorloc}
\end{equation} 
Following \cite{Booth:2005qc,Figueras:2009iu} we associate the non-equilibrium  entropy density $s$
of the boundary QFT  with the Bekenstein-Hawking entropy density of the apparent horizon  
\begin{equation}
a^3 s =\frac{\Sigma^3}{4 G_5}\bigg|_{r=r_{AH}}= \frac{N^2\Sigma^3}{16\pi}\bigg|_{r=r_{AH}}\,,
\eqlabel{as}
\end{equation}
with $a$ being the QFT background geometry scale factor \eqref{desitter1}. 
Using the holographic background equations of motion \eqref{ev1}-\eqref{mom} 
we find \cite{Buchel:2016cbj}
\begin{equation}
\frac{d(a^3 s)}{dt}=\frac{2N^2}{\pi}\ (\Sigma^3)'\ \frac{
 (d_+\chi)^2+3 (d_+\a)^2)}{-4 \calp}\bigg|_{r=r_{AH}}\,.
\eqlabel{dasdt}
\end{equation}
It was shown in \cite{Buchel:2016cbj} that \eqref{dasdt} correctly reproduces the 
entropy growth in the hydrodynamic regime, due to the bulk viscosity of the gauge theory plasma. 
In Appendix \ref{th1} we prove that the entropy production rate as defined by \eqref{dasdt}
is non-negative, \ie 
\begin{equation}
\frac{d(a^3 s)}{dt}\ge 0
\eqlabel{dasdt2}
\end{equation}
in holographic dynamics governed by \eqref{ev1}-\eqref{mom},

Dynamical evolution of homogeneous and isotropic states of $\caln=2^*$ gauge theory in $dS_4$
within  holographic framework proceeds as follows:
\begin{itemize}
\item First, we define parameters of the theory providing the mass parameters $\{m_b,m_f\}$
and the Hubble constant $H$. 
\item An initial state is specified providing gravitational scalar profiles 
\begin{equation}
\a(t=0,r)=\a_{initial}(r)\,,\qquad \chi(t=0,r)=\chi_{initial}(r)\,,
\eqlabel{alchiinitial}
\end{equation} 
subject to the boundary conditions \eqref{bcdata}.
The Hamiltonian constraint \eqref{ham} and the last two evolution equations in \eqref{ev1}
subject to the boundary conditions \eqref{bcdata} along with 
\begin{equation}
d_+\a=\frac{2m_b^2\ln r}{3r}+\calo(r^{-1})\,,\qquad d_+\chi=-\frac {m_f}{4}+\calo(r^{-1})\,,
\eqlabel{dpbc}
\end{equation}
are solved to determine $\Sigma_{initial}$ and $\{\a,\chi\}_{initial}$,
\begin{equation}
\Sigma_{initial}(r)\equiv \Sigma(t=0,r)\,,\qquad \{\a,\chi\}_{initial}(r)\equiv\{\a,\chi\}(t=0,r)\,.
\eqlabel{salchiini}                     
\end{equation}
Initial state specification is completed with solving for $A_{initial}(r)\equiv A(t=0,r)$ (the second equation in \eqref{ev1})
subject to the boundary condition \eqref{extension} with $\a_{2,0}(t=0)$ extracted from $\a_{initial}$ and specifying 
a free parameter $a_{2,0}^{initial}$, $a_{2,0}(t=0)=a_{2,0}^{initial}$. The latter parameter determines the 
initial energy density of the state, following \eqref{vev1}.  
\item This initial state can be evolved following  \cite{Chesler:2013lia}. At any time $t$ we can compute the one point correlation 
functions \eqref{vev1}-\eqref{vev4} and the entropy density of the state \eqref{as}. It is also possible to compute additional 
non-local observables of the evolution, for example as in \cite{Buchel:2014gta}. 
\end{itemize}

Implementing the holographic evolution as explained above is technically challenging, and is beyond the scope of this paper.
The difficulty\footnote{It is related to the presence of log-terms in 
the near-boundary expansion for the dual gravitational scalars \eqref{extension}. We hope to solve this problem in the 
future.} is not intrinsic to a $dS_4$ background metric for the boundary gauge theory --- in fact,
we prove in Appendix \ref{th2} the precise map between the holographic evolution of the $\caln=2^*$ states in de Sitter 
and in Minkowski space-times:

\bigskip
\begin{center}
 \begin{tabular}{|c |c |c|} 
 \hline
  & $dS_4$ & $R^{3,1}$  \\ 
 \hline
 background metric & $ds^2=-dt^2+e^{2Ht}  d\boldsymbol{x}^2$ &   $d{\hat{s}}^2=-d\tau^2+ d\boldsymbol{x}^2$ \\ 
 \hline
 evolution time & $t\in [0,\infty)$ &   $\t=(1-e^{-H t})/H\ \in \left[0,\frac 1H\right)$ \\ 
 \hline
 mass parameters & $\{m_b,m_f\}$ &   $\{\hat{m}_b(\t)=\frac{m_b}{1-H\t},\hat{m}_f(\t)=\frac{m_f}{1-H\t}\}$ \\ 
 \hline
 \end{tabular}\par
\bigskip
\end{center}
\bigskip

\section{Vacuum of $\caln=2^*$ gauge theory in $dS_4$}\label{venergy}

In the previous section we outlined the setup for the holographic evolution of a generic homogeneous and 
isotropic state of $\caln=2^*$ gauge theory. While we have not implemented the evolution, it is physically 
reasonable to expect that any initial, finite energy density state of the theory in de Sitter background  would evolve 
as $t\to \infty$ into 
a universal static state,  which we label as {\it vacuum}.  Indeed, one expects that energy density $\cale(t)$ is 
diluted according to \eqref{cons}, until it reaches the (constant) vacuum value $\cale_v$, entirely determined 
by the Hubble constant $H$ and the relevant microscopic couplings of the theory $\{m_b, m_f\}$, \ie $\cale_v=\cale_v(H,m_b,m_f)$:
\begin{equation}
\lim_{t\to \infty}\cale(t)=-\lim_{t\to \infty} P(t)= \cale_v\,,\qquad \lim_{t\to \infty}\dot\cale(t)=0\,.
\eqlabel{dil1}
\end{equation}
From \eqref{vev1} we then expect that the normalizable coefficients
of the gravitational bulk dynamics $\{a_{2,0},\chi_{3,0},\a_{2,0}\}$ approach
constant values at late times:
\begin{equation}
\begin{split}
&\lim_{t\to \infty}\{a_{2,0},\chi_{3,0},\a_{2,0}\}(t)=\{a_{2,0},\chi_{3,0},\a_{2,0}\}^v\,,\\
&\lim_{t\to \infty}\{\dot{a}_{2,0},\dot{\chi}_{3,0},\dot{\a}_{2,0}\}(t)=\{0,0,0\}\,.
\end{split}
\eqlabel{latenom}
\end{equation}
Note that $a_{2,0}^v$ is not independent from $\{\chi_{3,0},\a_{2,0}\}^v$;
from \eqref{ward2}:
\begin{equation}
a_{2,0}^v=\frac{152}{9} H^2 m_f^2-\frac{104}{243} m_b^4
+\frac{56}{27} m_f^4+\frac{4}{27} \alpha_{2,0}^v m_b^2-\frac13
\chi_{3,0}^v m_f-\frac13 (\alpha_{2,0}^v)^2\,.
\eqlabel{a20vres}
\end{equation}
Given \eqref{latenom}, it is straightforward to see that to any order
in the boundary expansion $\{\a,\chi,A\}$ become static at late times:
\begin{equation}
\lim_{t\to \infty}\{\a,\chi,A\}(t,r)=\{\a,\chi,A\}_v(r)\,.
\eqlabel{lateachiA}
\end{equation}
Finally, \eqref{ham} along with the boundary condition \eqref{bcdata}
implies that $\frac{\Sigma}{a}$ becomes static at late times:
\begin{equation}
\lim_{t\to \infty} \frac{\Sigma(t,r)}{a(t)}=\sigma_v(r)\,.
\eqlabel{slate}
\end{equation}

Dropping the time dependence according to \eqref{lateachiA} and \eqref{slate}
we obtain from \eqref{ev1}-\eqref{mom} the following set of dual
gravitational equations describing the vacuum of $\caln=2^*$ gauge theory
in $dS_4$:
\begin{equation}
\begin{split}
&0=\a_v''+\left(\frac{3H}{2A_v}+\left(\ln A_v\sigma_v^3\right)'\right)\a_v'-\frac{1}{48 A_v}\del_\a\calp\,,\\
&0=\chi_v''+\left(\frac{3H}{2A_v}+\left(\ln A_v\sigma_v^3\right)'\right)\chi_v'-\frac{1}{16 A_v}\del_\chi\calp\,,\\
&0=\sigma_v''+\frac 43 \sigma_v\left(3(\a_v')^2+(\chi_v')^2\right)\,,\\
&0=A_v''+4 A_v'\left(3(\a_v')^2+(\chi_v')^2\right) -6 A_v \left((\ln \sigma_v)'\right)^2-6 H(\ln\sigma_v)'-\frac \calp6\,,
\end{split}
\eqlabel{veoms}
\end{equation}
along with the constraints
\begin{equation}
\begin{split}
&0=\sigma_v'+\frac{\sigma_v}{2A_v}(H-A_v')\,,\\
&0=(\a_v')^2+\frac 13 (\chi_v')^2-\frac 12 \left((\ln\sigma_v)'\right)^2-\frac{\sigma_v'}{4\sigma_vA_v}(2H+A_v')+
\frac{H}{8A_v^2}(H-A_v')-\frac{\calp}{24A_v}\,.
\end{split}
\eqlabel{veoms2}
\end{equation}
It is straightforward to verify that constraints \eqref{veoms2} are consistent with \eqref{veoms}.

Vacuum solution has to satisfy the boundary condition (see \eqref{bcdata})
\begin{equation}
\begin{split}
&\sigma_v=\frac {r+\l}{2}+\calo(r^{-1})\,,\qquad A_v=\frac{(r+\l)^2}{8}-H (r+\l) +\calo(r^0)\,,\\
&\a_v=-\frac{8m_b^2\ln r}{3r^2}+\calo(r^{-2})\,,\qquad \chi_v=\frac{2m_f}{r}+\calo(r^{-2})\,,
\end{split}
\eqlabel{bclatet}
\end{equation}
where we reintroduced time-independent radial coordinate shift symmetry $r\to r+\l$, see \cite{Chesler:2013lia}.
It is important to realize that the boundary condition \eqref{bclatet} is not sufficient to identify 
a unique solution. Note that at late times the apparent horizon settles at a fixed radial location $r_{AH}$ 
determined from 
\begin{equation}
d_+\Sigma(t,r_{AH})=0\qquad \Longleftrightarrow\qquad  H \sigma_v+A_v \sigma'_v\bigg|_{r=r_{AH}}=0 \,.
\eqlabel{AHlate}
\end{equation} 
We need to construct a solution to \eqref{veoms}-\eqref{veoms2} for $r\in( r_{AH},\infty)$.
As shown in Appendix \ref{th1}, $\sigma'_v>0$ for $r>r_{AH}$; additionally, the absence of naked singularities in the 
gravitational dual requires that $\sigma_v>0$ for $r>r_{AH}$ --- thus, \eqref{AHlate} implies that $A_v(r_{AH})<0$. 
Since $A_v\to \infty$ as $r\to \infty$, there must be a radial location $r=r_s > r_{AH}$, such that $A_v(r_s)=0$. 
Inspection of the bulk  equations of motion \eqref{veoms}-\eqref{veoms2} shows that they are generically singular 
when $A_v=0$. Avoiding this singularity  provides complementary conditions to uniquely identify the $dS_4$ vacuum
of $\caln=2^*$ gauge theory.    

The rest of this section is organized as follows. We prove that the $dS_4$ vacuum of $\caln=2^*$ gauge theory 
obtained as a late-time attractor of the evolution from a homogeneous and isotropic initial state is a 
Bunch-Davies vacuum. We explicitly compute the vacuum energy of the theory $\cale_v$ for $\{m_f=0,m_b\ne 0\}$ and in a 
 supersymmetric case $m_f=m_b$. Following \eqref{dasdt}, we show that non-conformal gauge theory 
vacuum is not an adiabatic state --- the rate of the comoving entropy production 
is non-zero:
\begin{equation}
\lim_{t\to \infty}\ \frac{1}{H^3 a^3}\frac{d}{dt} (a^3 s)\equiv 3 H\times \calr  ={\rm const}\ne 0\,.
\eqlabel{dsdtlate}
\end{equation}

To contrast dynamics of non-conformal gauge theories with conformal ones, we review 
$\caln=4$ SYM in $dS_4$ in Appendix \ref{n4ds}.   

\subsection{$dS_4$ vacuum is Euclidean}

Bunch-Davies (Euclidean) vacuum of $\caln=2^*$ gauge theory was constructed in \cite{Buchel:2003qm}. 
We review the main points of the construction and prove that the holographic vacuum equations \eqref{veoms}
and \eqref{veoms2} are equivalent to those in \cite{Buchel:2003qm}.

To describe Bunch-Davies vacuum of $\caln=2^*$ gauge theory at strong coupling one studies the effective gravitational action 
\eqref{action5} within the following background ansatz:
\begin{equation}
ds_5^2=-c_1^2\ (d\calm_4)^2 +c_2^2\ (dz)^2\,,\qquad c_i=c_i(z)\,,\qquad  \chi=\chi(z)\,,\qquad \a=\a(z)\,,
\eqlabel{FGvacuum}
\end{equation} 
where $(d\calm_4)^2$ is the de Sitter boundary gauge theory metric. Irrespectively whether this  metric is written 
in flat or closed slicing $(d\calm_4^{f/c})^2$,
\begin{equation}
(d\calm_4^{f})^2=-dt^2 + e^{2 Ht} d\boldsymbol{x}^2\,,\qquad (d\calm_4^{c})^2=-dt^2 + \frac{1}{H^2}\ \cosh^2(Ht)\ (dS^3)^2 \,,
\eqlabel{ocslice}
\end{equation}
the dual gravitational equations of motion take form 
\begin{equation}
\begin{split}
&0=\a''+\left(\ln\frac{c_1^4}{c_2}\right)' \a' -\frac{c_2^2}{24}\ \del_\a \calp\,,\\
&0=\chi''+\left(\ln\frac{c_1^4}{c_2}\right)' \chi' -\frac{c_2^2}{8}\ \del_\chi \calp\,,\\
&0=c_1''+\frac{(c_1')^2}{c_1}-\frac{c_1'c_2'}{c_2}-\frac{c_2^2H^2}{c_1}
+2c_1\left({(\a')^2}+\frac 13 (\c')^2\right)+\frac 16c_1c_2^2\ \calp\,,
\\
&0={(\a')^2}+\frac 13 (\c')^2-\frac{(c_1')^2}{c_1^2}+\frac{c_2^2H^2}{c_1^2}-\frac{1}{12}c_2^2\ \calp\,.
\end{split}
\eqlabel{fgeoms}
\end{equation}
Notice that closed slicing de Sitter metric makes Euclidean analytic continuation obvious: 
\begin{equation}
(d\calm_4^{c})^2\bigg|_{t\to it\equiv t_E\equiv \frac{\theta}{H}}\qquad \to\qquad \frac{1}{H^2}\ 
\left(d\theta^2+\cos^2\theta\ (dS^3)^2\right)=\frac{1}{H^2}\ (dS^4)^2\,.
\eqlabel{eucl}
\end{equation} 

It is straightforward to relate \eqref{fgeoms} and \eqref{veoms},
\eqref{veoms2}: assuming that as $z\to \infty$ the warp factor $c_1\to \infty$, the systems 
of equations become identical with
\begin{equation}
\begin{split}
&r=-\int_z^\infty ds\ c_1(s)c_2(s)\,,\qquad \sigma_v(r)=c_1(z)\ \exp\left[
H \int_z^\infty ds\ \frac{c_2(s)}{c_1(s)}\right]\,,\\
& A_v(r)=\frac{c_1(z)^2}{2}\,,\qquad \a_v(r)=\ln\r(z)\,,\qquad \chi_v(r)=\chi(z) \,.
\end{split}
\eqlabel{mapfgef}
\end{equation}
Of course, the map \eqref{mapfgef} is nothing but the map between the asymptotic $t\to \infty $ infalling 
Eddington-Finkelstein metric \eqref{EFmetric} and the Fefferman-Graham metric  \eqref{FGvacuum}.  
The advantage to use the characteristic formulation in the dual gravitational description  \eqref{EFmetric}
is that it clarifies the  selection of the Bunch-Davies  vacuum for the strongly coupled gauge theory. 
As we will see shortly, it also explains why the Bunch-Davies vacuum is non-adiabatic.

\subsection{$\cale_v$ of $\caln=2^*$ gauge theory in $dS_4$}

In this section we numerically solve the equations of motion \eqref{veoms}, \eqref{veoms2}  and compute $\cale_v$ 
for two representative holographic RG flows: $m_f=m_b$ and $\{m_f=0, m_b\ne 0\}$.

Introducing a new radial coordinate 
\begin{equation}
x=\frac{H}{r}\,,
\eqlabel{defx}
\end{equation}
and redefining  
\begin{equation}
A_v\equiv\frac{H^2}{8x^2} a_0(x)\,,\qquad \sigma_v\equiv \frac{H}{2x}\ s_0(x)\,,\qquad \a_v\equiv \ln(r_0(x))\,,\qquad \cosh\chi_v\equiv c_0(x) \,,
\eqlabel{wardx}
\end{equation}
from \eqref{veoms} and \eqref{veoms2} we find:
\begin{equation}
\begin{split}
&0=r_0''+\frac{2r_0'}{x}-\frac{(r_0')^2}{r_0} -\frac{5r_0'\Theta}{3a_0 x r_0^2}  -\frac23 \frac{2 c_0^2 r_0^{12} (c_0^2-1)-r_0^6 (2 c_0^2-1)+1}{x^2 r_0^3 a_0}\,,
\end{split}
\eqlabel{xeoms1}
\end{equation}
\begin{equation}
\begin{split}
&0=c_0''+\frac{2 c_0'}{x}-\frac{c_0 (c_0')^2}{c_0^2-1} -\frac{5c_0'\Theta}{3a_0 x r_0^2}-(c_0^2-1) \frac{((2 c_0^2-1) r_0^6-4) r_0^2 c_0}{a_0 x^2}\,,
\end{split}
\eqlabel{xeoms2}
\end{equation}
\begin{equation}
\begin{split}
&0=a_0'-\frac{2 a_0}{x}+\frac{2\Theta}{3r_0^2 x} \,,
\end{split}
\eqlabel{xeoms3}
\end{equation}
\begin{equation}
\begin{split}
&0=\frac{s_0'}{s_0}-\frac1x-\frac{4}{a_0}+\frac{\Theta}{3a_0xr_0^2} \,,
\end{split}
\eqlabel{xeoms4}
\end{equation}
where 
\begin{equation}
\begin{split}
\Theta^2=9 (r_0')^2 r_0^2 a_0^2 x^2+\frac{3 a_0^2 x^2 r_0^4 (c_0')^2}{c_0^2-1}+(3 c_0^2 r_0^{12} (1-c_0^2)+6 r_0^6 (2 c_0^2-1)+3) a_0+144 x^2 r_0^4\,.
\end{split}
\eqlabel{defTHETA}
\end{equation}
Equations \eqref{xeoms1}-\eqref{xeoms4} have the following boundary asymptotics (we keep only the couple relevant terms)
\begin{equation}
\begin{split}
&c_0=1+x^2 c_{2,0}-x^3 c_{2,0} a_{1,0}+x^4 \left(
\left(\frac{16}{3} c_{2,0}^2+32 c_{2,0}\right) \ln x+c_{4,0}
\right)+\calo(x^5\ln x)\,,\\
&r_0=1+x^2 \left(r_{2,1} \ln x+r_{2,0}\right)+\calo(x^3\ln x)
\,,\qquad a_0=1+a_{1,0} x+\calo(x^2)\,,\\
&s_0=1+\left(4+\frac{a_{1,0}}{2}\right)x+\calo(x^2)\,,
\end{split}
\eqlabel{uvx}
\end{equation}
where the non-normalizable coefficients $\{r_{2,1},c_{2,0}\}$ and the gauge parameter $a_{1,0}$ are related to 
$\{m_b,m_f\}$ and $\l$ in \eqref{bclatet} as follows
\begin{equation}
r_{2,1}=\frac {8m_b^2}{3H^2}\,,\qquad c_{2,0}=\frac{2m_f^2}{H^2}\,,\qquad a_{1,0}=\frac{8\l}{H}\,.
\eqlabel{mapnonnom}
\end{equation}
The normalizable coefficients $\{r_{2,0},c_{4,0}\}$ determine expectation values $\{\calo_2^v,\calo_3^v\}$ and the vacuum energy density $\cale_v$.
Specifically, 
\begin{equation}
\begin{split}
\frac{32\pi^2}{N^2}\ \cale_v&=3 H^4+\frac{r_{2,0}}{2}  m_b^2 H^2-\frac{c_{4,0}}{16}  H
^4+\frac{3(a_{1,0})^2}{32}  m_f^2 H^2
\\&+ \biggl(\frac43 (m_b^4-m_f^4)-4 m_f^2 H^2\biggr)\ln\frac{H}{\Lambda}
+q_1 m_f^2 H^2+q_2 m_f^4+q_3 m_b^2 H^2+q_4 m_b^4\,,
\end{split}
\eqlabel{caleres}
\end{equation}  
where $\Lambda$ and $q_i$ are arbitrary renormalization scheme choice parameters, 
related to $\dd_i$ in \eqref{vevsamb}. The ambiguities of the vacuum energy 
density discussed here mirror the well-known ambiguities defining the 
stress-energy tensor of weakly coupled QFTs in curved space-times \cite{Birrell:1982ix}.
In section \ref{conclude} we discuss the implication of these ambiguities 
in the context of self-consistent inflationary scenarios \eqref{desitter2}. 

As we already mentioned, the choice of the gauge parameter $\l$ (correspondingly $a_{1,0}$ in \eqref{uvx}) is completely 
arbitrary. We find it convenient to adjust $a_{1,0}=a_{1,0}(r_{2,1},c_{2,0})$ in such a way that the zero of the 
warp function $a_0$ occurs at exactly the same location as it is for the $\caln=4$ SYM duality where $r_{2,1}=c_{2,0}=0$
(see \eqref{defxs}):
\begin{equation}
a_0(x)\bigg|_{x=x_s=\frac 19} = 0\,,
\eqlabel{inforcexs}
\end{equation} 
implying that the range of $x$ is 
\begin{equation}
x\in \left(0,\frac 19\right]\,.
\eqlabel{ranges}
\end{equation}
The absence of singularities in \eqref{xeoms1}-\eqref{xeoms4} as $y\equiv \left(\frac 19-x\right)\to 0_+$ enforces the 
following asymptotic behaviour
\begin{equation}
\begin{split}
&a_0=8 y+\left(\frac{27}{r_{0,s,0}^4}-27 r_{0,s,0}^2 c_{0,s,0}^2 (c_{0,s,0}^2r_{0,s,0}^6-r_{0,s,0}^6-4)-54r_{0,s,0}^2-72\right)y^2+\calo(y^3)\,,\\
&s_0=s_{0,s,0}+\calo(y)\,,\qquad r_0=r_{0,s,0}+\calo(y)\,,\qquad c_0=c_{0,s,0}+\calo(y)\,,\qquad 
\end{split}
\eqlabel{irass}
\end{equation}
determined by 3 parameters $\{r_{0,s,0},c_{0,s,0},s_{0,s,0}\}$. The total order of the system of the differential equations 
\eqref{xeoms1}-\eqref{xeoms4} is 6; thus, along with the normalizable coefficients $\{r_{2,0}, c_{4,0}\}$ 
and the gauge parameter $a_{1,0}$, 
we have the correct number of the adjustable coefficients necessary to uniquely identify the gravitational solution 
for a given set of $\left\{\frac{m_b}{H},\frac{m_f}{H}\right\}$. 

In the limit $\{r_{2,1},c_{2,0}\}\ll 1$ \eqref{xeoms1}-\eqref{xeoms4} can be solved analytically. To order $\calo(r_{2,1}^2,c_{2,0})$,
$a_0$ and $s_0$ are that of the $\caln=4$ SYM, discussed in Appendix \ref{n4ds}, while
\begin{equation}
\begin{split}
r_0=&1-r_{2,1}\
 x^2 \left(
\frac{5x-1}{2}\ \ln  \frac{\sqrt{(1-9 x) (1-x)}+5 x-1}{5 x-1-\sqrt{(1-9 x) (1-x)}} -\sqrt{(1-9 x) (1-x)}\right)\\
&\times  \frac{1}{(1-9 x)^{3/2} (1-x)^{3/2}}
+\calo(r_{2,1}^2,c_{2,0}^2)\,,
\end{split}
\eqlabel{r0pert}
\end{equation}    
\begin{equation}
\begin{split}
c_0=&1+c_{2,0}\ x^2 \left(8x^2\ \ln \frac{\sqrt{(1-9 x) (1-x)}+5 x-1}{5 x-1-\sqrt{(1-9 x) (1-x)}}-(5x-1)\sqrt{(1-9 x) (1-x)}\right)^2\\
&\times  \frac{1}{(1-9 x)^3 (1-x)^3}+\calo(r_{2,1}^2,c_{2,0}^2)\,.
\end{split}
\eqlabel{c0pert}
\end{equation}    
For general values of $\{r_{2,1},c_{2,0}\}$ the system \eqref{xeoms1}-\eqref{xeoms4} is solved numerically, using the shooting 
method introduced in \cite{Aharony:2007vg}.

\begin{figure}[t]
\begin{center}
\psfrag{x}{{$x$}}
\psfrag{y}{{$r_0$}}
\psfrag{t}{{$c_0$}}
\psfrag{g}{{$\frac{m_b^2}{H^2}=0.15$}}
\psfrag{h}{{$\frac{m_b^2}{H^2}=\frac{m_f^2}{H^2}=0.15$}}
  \includegraphics[width=2.5in]{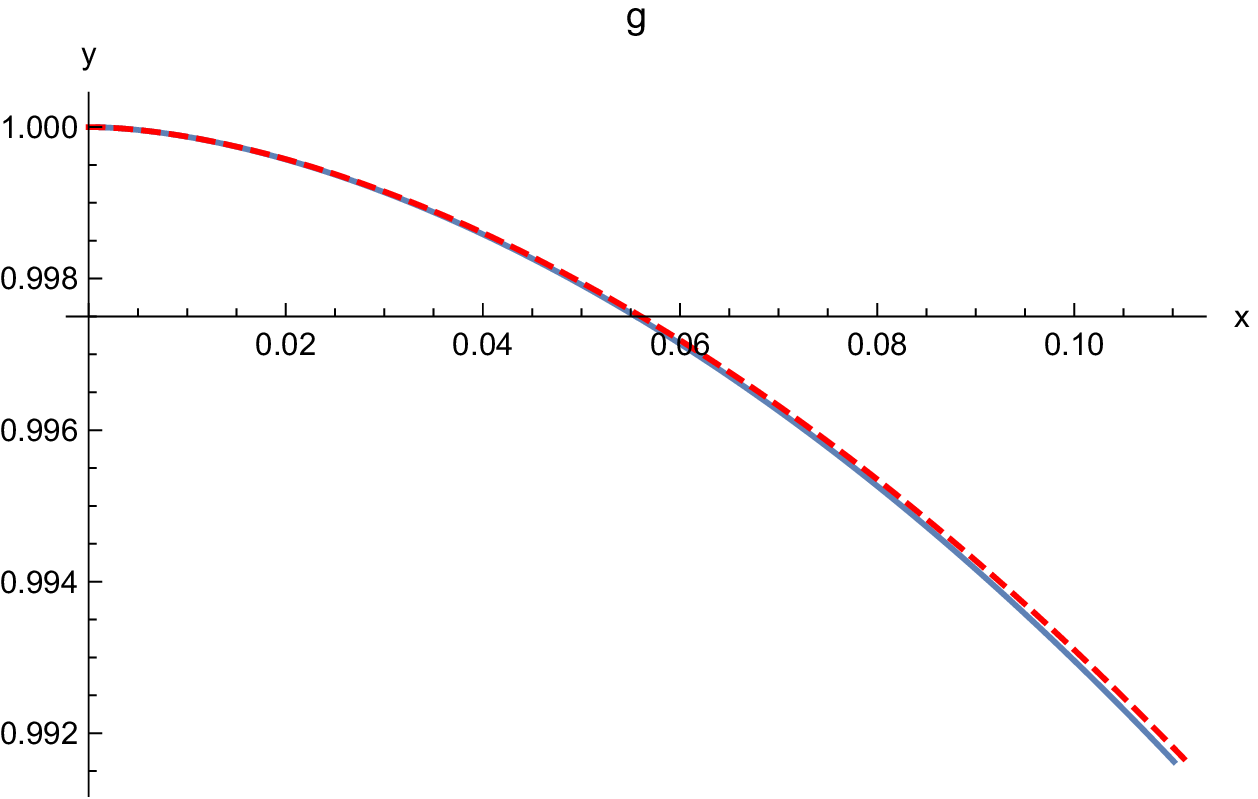}\,\,\,\,\,\,\,\,\,\,\,\,
  \includegraphics[width=2.5in]{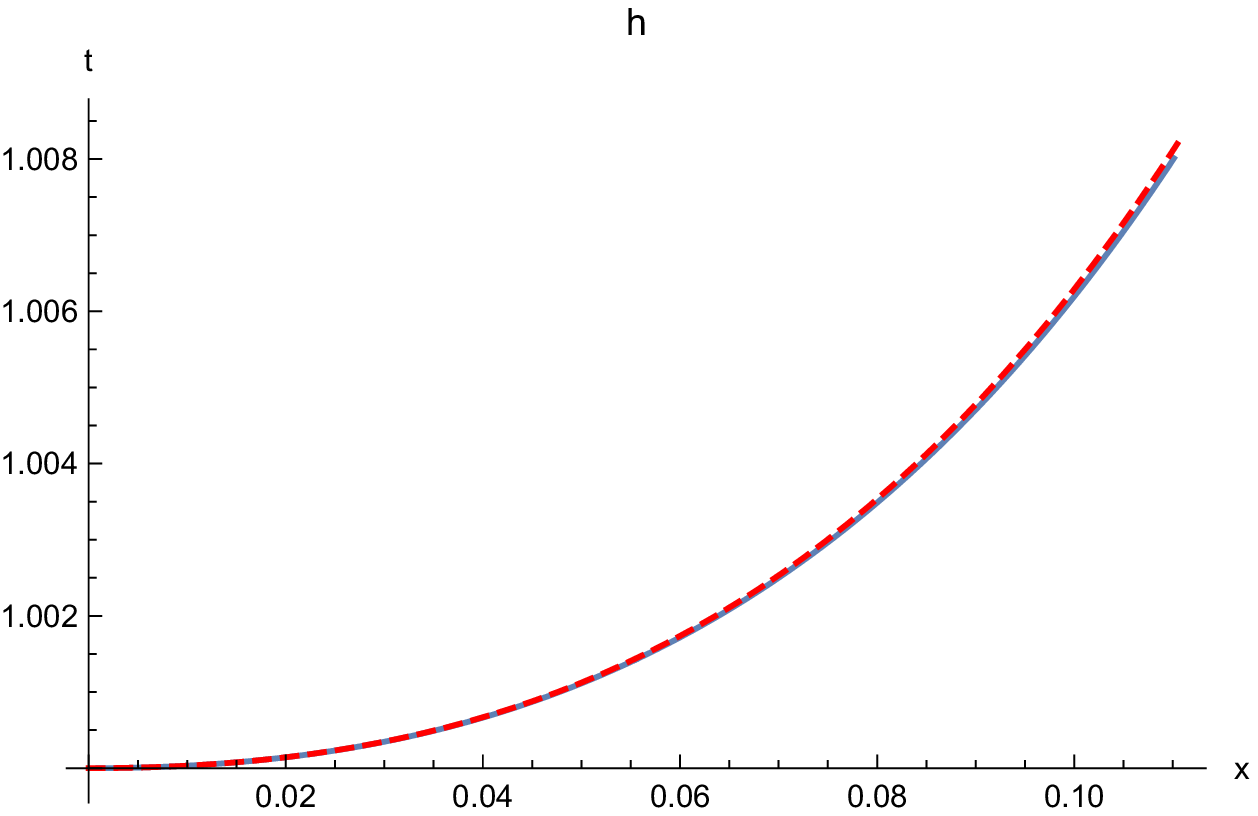}
\end{center}
 \caption{Profiles of the gravitational scalars $\{r_0,c_0\}$ (solid curves) describing the holographic dual to 
$\caln=2^*$ gauge theory vacuum in $dS_4$ for select values of $\{m_b,m_f\}$. 
The dashed red lines represent the analytic solutions valid for $\{m_b,m_f\}\ll H$, see
\eqref{r0pert} and \eqref{c0pert}.   
}\label{figure1a}
\end{figure}

Fig.\ref{figure1a} presents  characteristic profiles of the gravitational scalar $\{r_{0},c_{0}\}$ for select (small) values of 
$\{m_b,m_f\}$. For comparison, we added perturbative predictions given by \eqref{r0pert} and \eqref{c0pert}.

\begin{figure}[t]
\begin{center}
\psfrag{z}{{$\Lambda=H$}}
\psfrag{w}{{$\Lambda=m_b$}}
\psfrag{x}{{$\frac{m_b^2}{H^2}$}}
\psfrag{y}{{$\frac{32\pi^2}{N^2 }\ \frac{\cale_v}{\Lambda^4}$}}
  \includegraphics[width=2.5in]{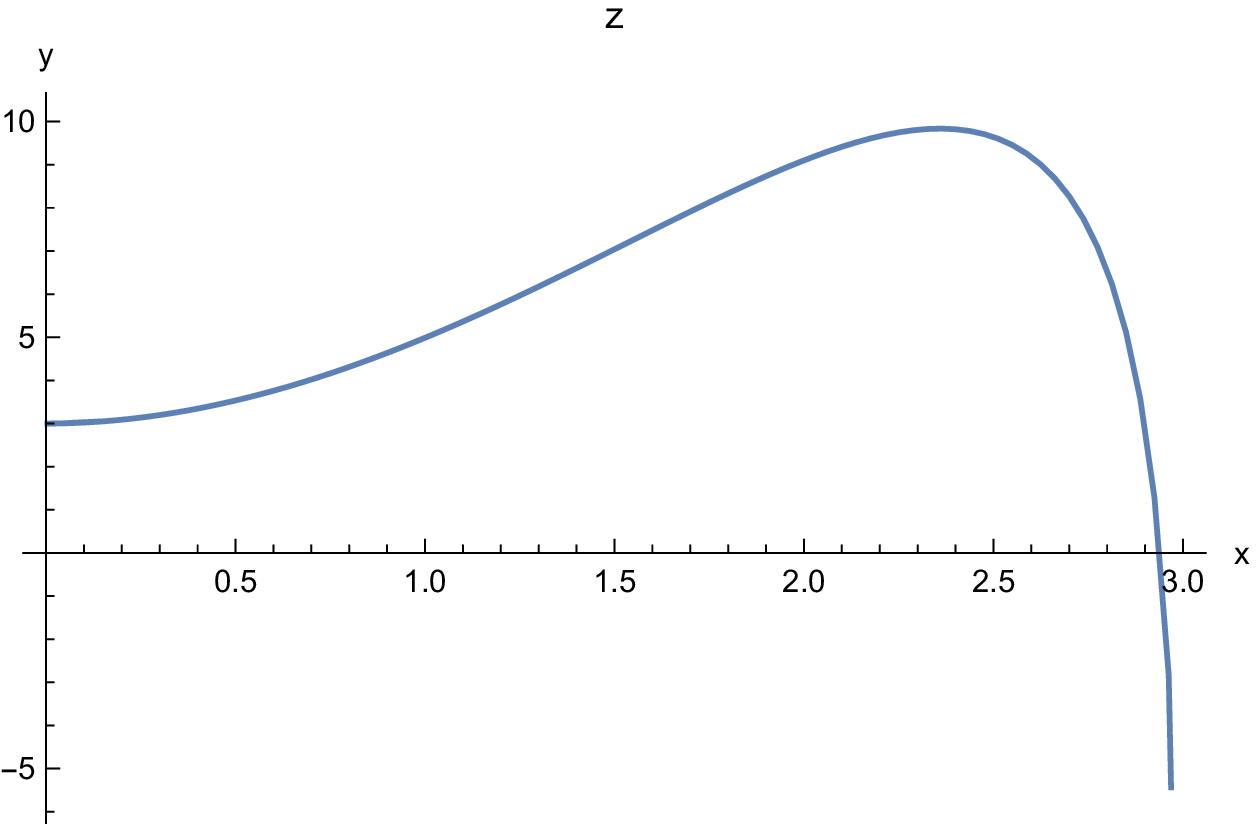}\,\,\,\,\,\,\,\,\,\,\,\,
  \includegraphics[width=2.5in]{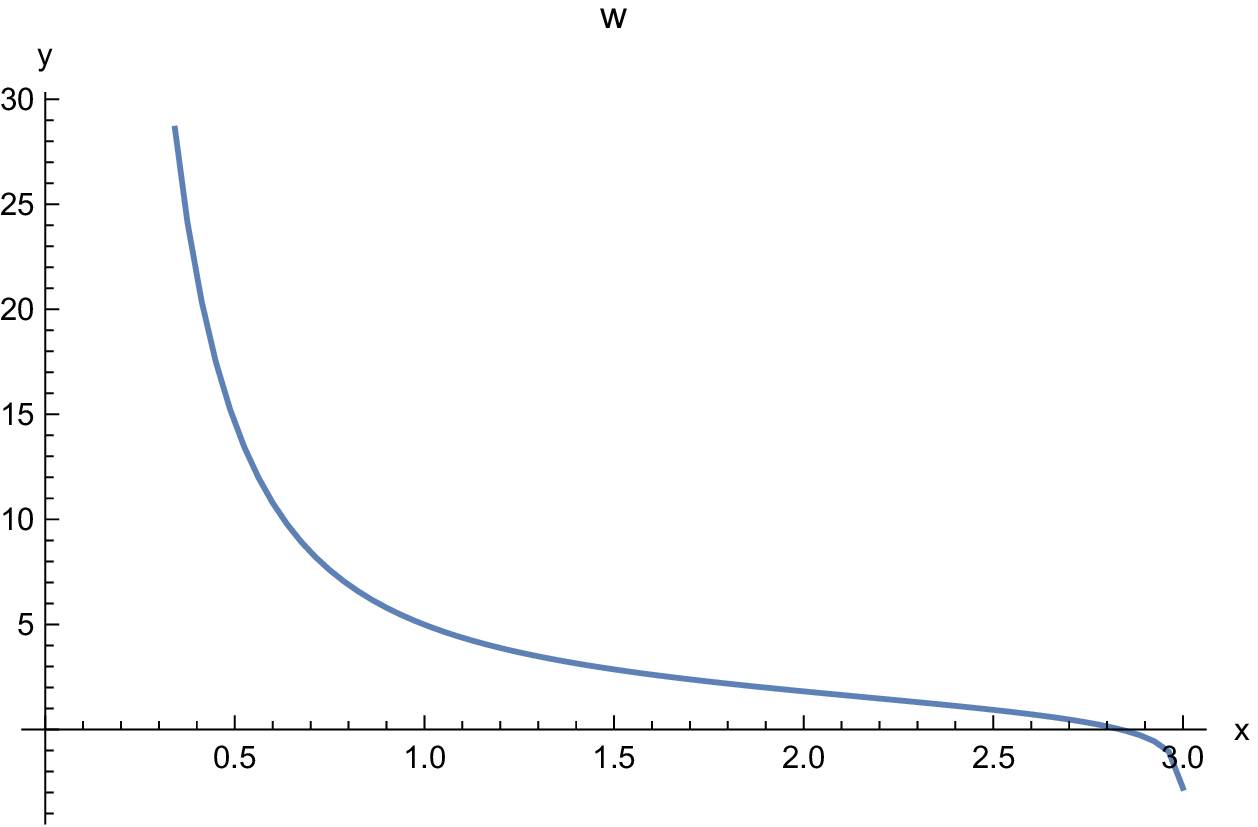}
\end{center}
 \caption{Vacuum energy $\cale_v$ (see \eqref{caleres}) of $\caln=2^*$ gauge theory in $dS_4$ with 
$\{m_b\ne 0, m_f=0\}$. We choose renormalization parameters $q_i=0$, and 
set the renormalization scale $\Lambda=H$ (left panel) and $\Lambda=m_b$ (right panel). }\label{figure3}
\end{figure}

\begin{figure}[t]
\begin{center}
\psfrag{z}{{$\Lambda=H$}}
\psfrag{w}{{$\Lambda=m_f=m_b$}}
\psfrag{x}{{$\frac{m_b^2}{H^2}$}}
\psfrag{y}{{$\frac{32\pi^2}{N^2 }\ \frac{\cale_v}{\Lambda^4}$}}
  \includegraphics[width=2.5in]{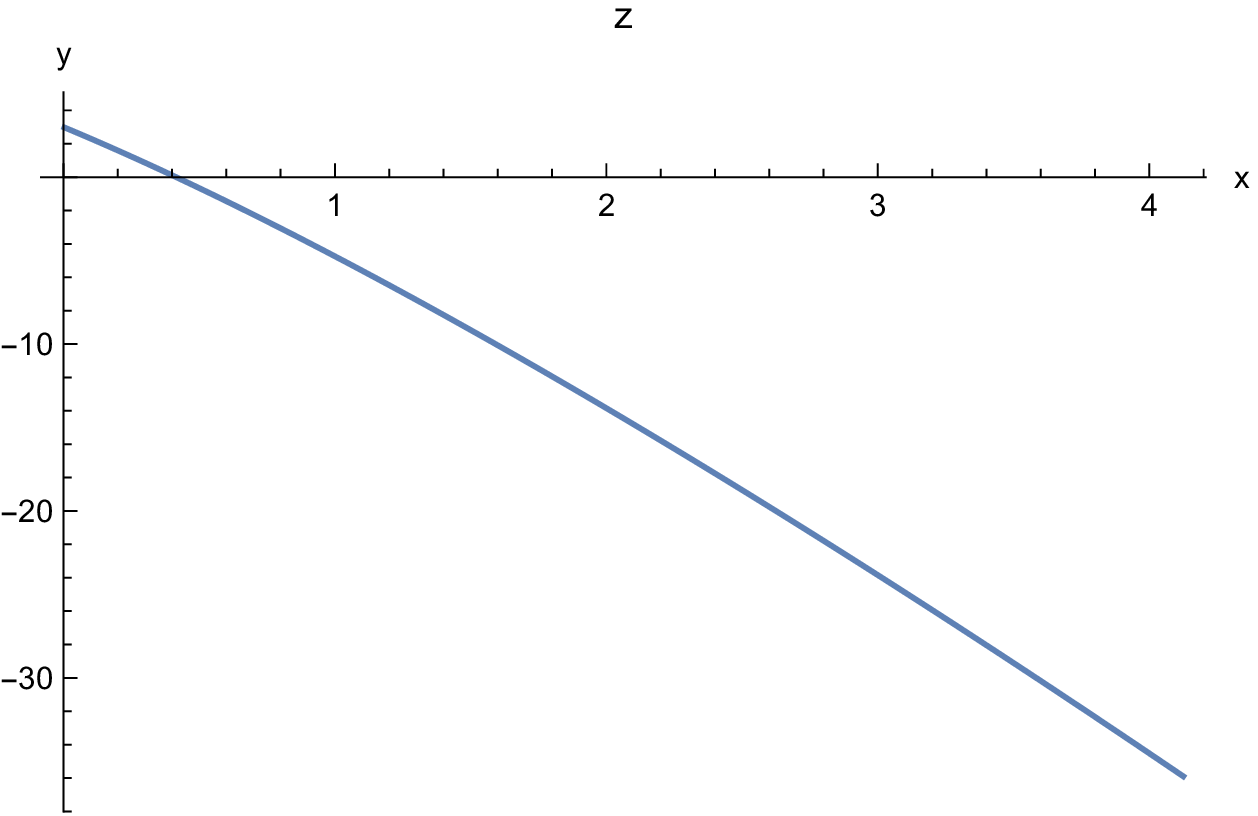}\,\,\,\,\,\,\,\,\,\,\,\,
  \includegraphics[width=2.5in]{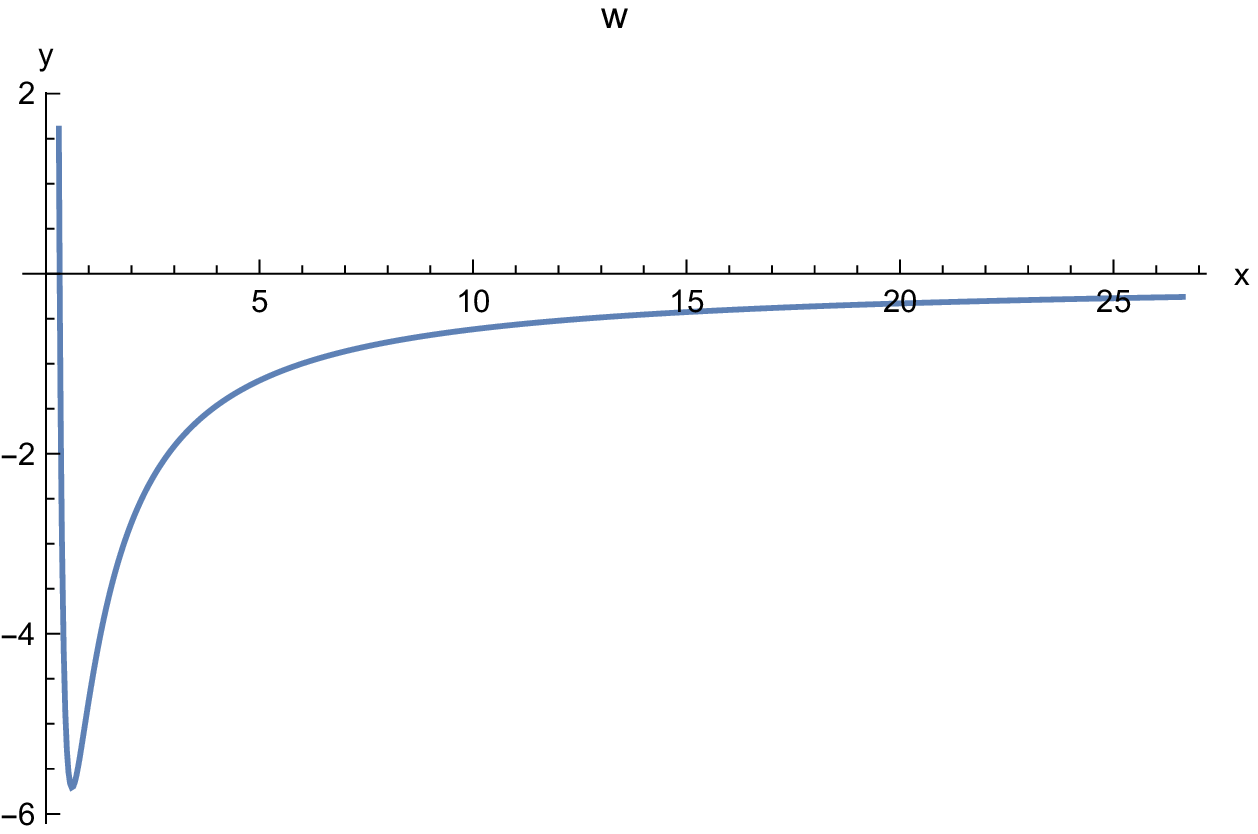}
\end{center}
 \caption{Vacuum energy $\cale_v$ (see \eqref{caleres}) of $\caln=2^*$ gauge theory in $dS_4$ with 
$\{m_b=m_f\}$. We choose renormalization parameters $q_i=0$, and 
set the renormalization scale $\Lambda=H$ (left panel) and $\Lambda=m_b$ (right panel).}\label{figure4}
\end{figure}

Having obtained numerical solutions to \eqref{xeoms1}-\eqref{xeoms4}, we extract the normalizable coefficients 
$\{r_{2,0},c_{4,0}\}$. We compute the vacuum energy density following \eqref{caleres}. The results are renormalization 
scheme dependent, and we set $q_i=0$. There is a remaining ambiguity associated with the choice of the 
renormalization scale $\Lambda$.  This ambiguity is important, because of the interesting $\ln \frac H\Lambda$ dependence in 
the expression for the vacuum energy. We consider setting $\Lambda=H$ and $\Lambda=m_b$. In the former case 
the log-dependence disappears, and one can take the limit $\frac{m_b}{H}\to 0$, reproducing the $\caln=4$ SYM 
result \eqref{n4ev}.  Results for the vacuum energy $\cale_v$ of $\caln=2^*$ gauge theory in $dS_4$ as a 
function of $\frac{m_b^2}{H^2}$ are collected in figs.~\ref{figure3}-\ref{figure4}. Modulo the scheme dependence, 
we find that $\frac{\cale_v}{\Lambda^4}>0$ for sufficiently small $\frac{m_b^2}{H^2}$; however, it becomes negative 
for large mass parameters $\{m_b,m_f\}$.

\subsection{$dS_4$ vacuum of a non-conformal gauge theory is non-adiabatic}

In Appendix \ref{n4ds} we argued that the comoving entropy of $\caln=4$ SYM vacuum in $dS_4$ is constant --- the vacuum 
is an adiabatic state. Here, we compute the comoving entropy production rate $\calr$ (see \eqref{dsdtlate}) of $\caln=2^*$
gauge theory and show that it is non-zero. Thus,  we are led to claim that  $dS_4$ vacua of non-conformal gauge theories 
are non-adiabatic.

\begin{figure}[t]
\begin{center}
\psfrag{x}{{$\left(x-\frac19\right)$}}
\psfrag{y}{{$0=\sigma_v'+\cdots$}}
\psfrag{z}{{$0=(\a_v')^2+\cdots$}}
  \includegraphics[width=2.5in]{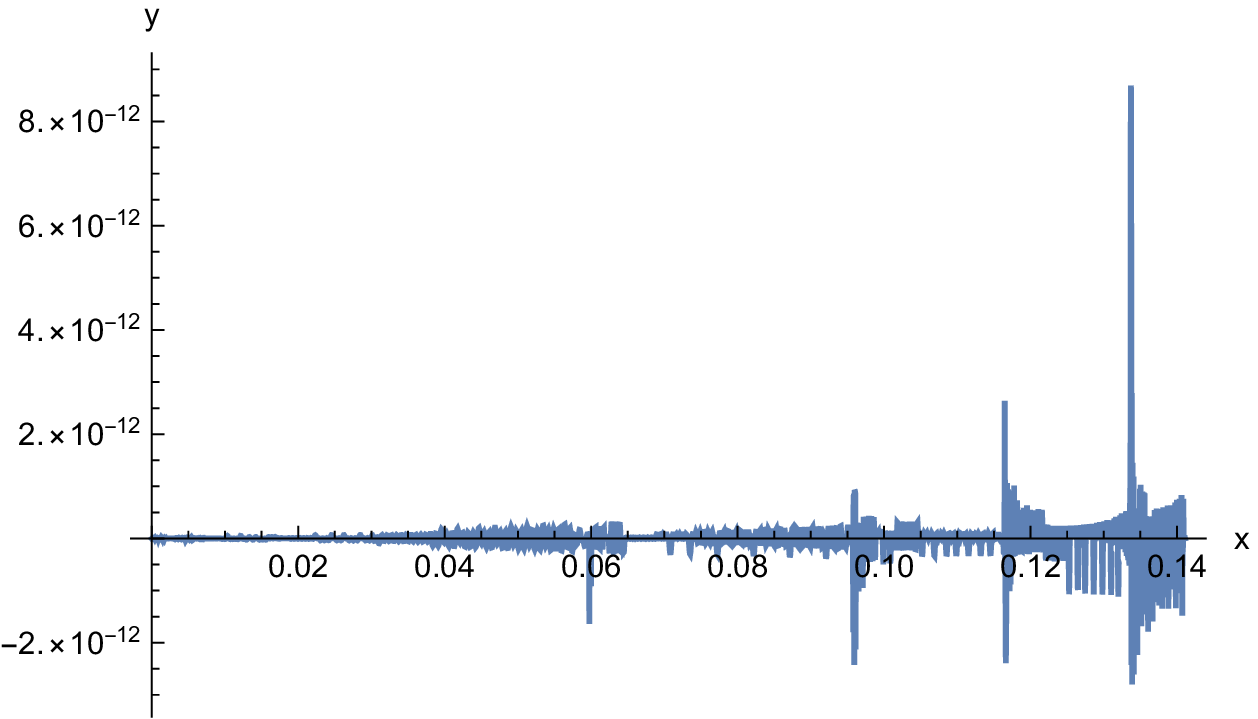}\,\,\,\,\,\,\,\,\,\,\,\,\,\,
  \includegraphics[width=2.5in]{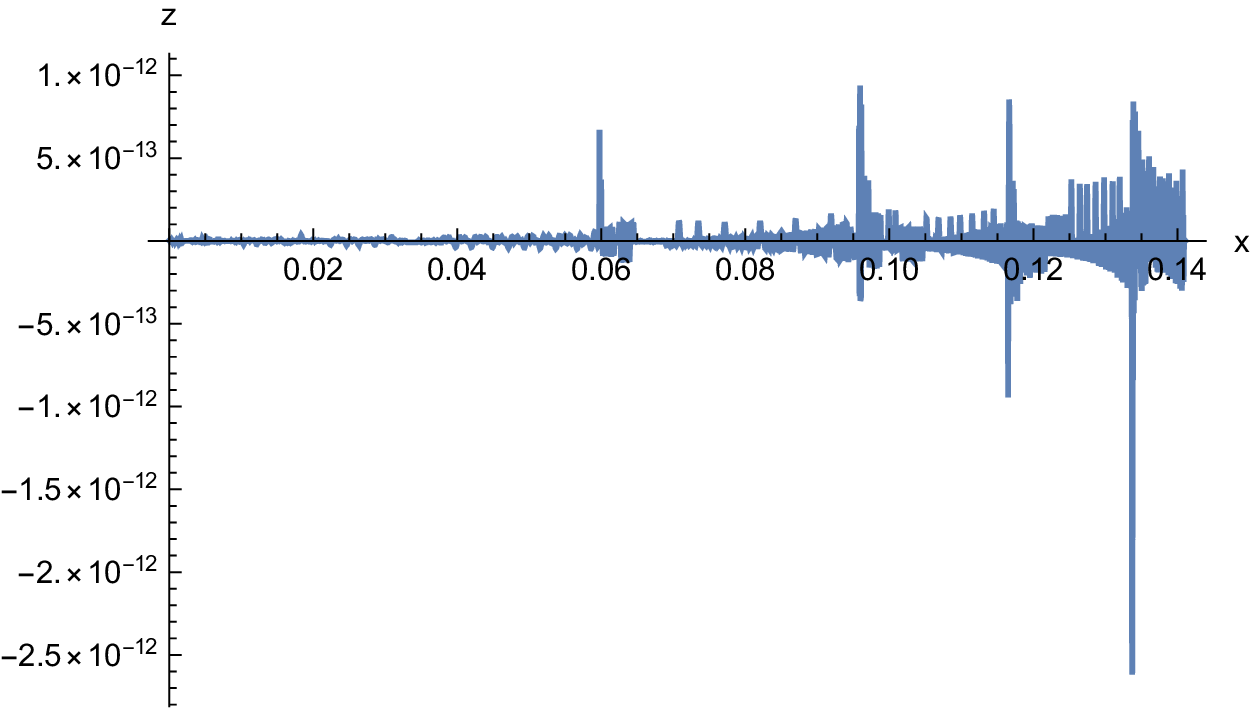}
\end{center}
 \caption{Constraint equations \eqref{veoms} are not integrated in constructing the gravitational background 
dual to vacuum of $\caln=2^*$ gauge theory in the vicinity of the apparent horizon, \ie for $x>x_s$. Rather, their residuals 
are used as a numerical test. Here $m_b=H$ and $m_f=0$.}\label{figure2}
\end{figure}

\begin{figure}[t]
\begin{center}
\psfrag{x}{{$\left(x-\frac 19\right)$}}
\psfrag{y}{{$0=\sigma_v'+\cdots$}}
\psfrag{z}{{$0=(\a_v')^2+\cdots$}}
  \includegraphics[width=2.5in]{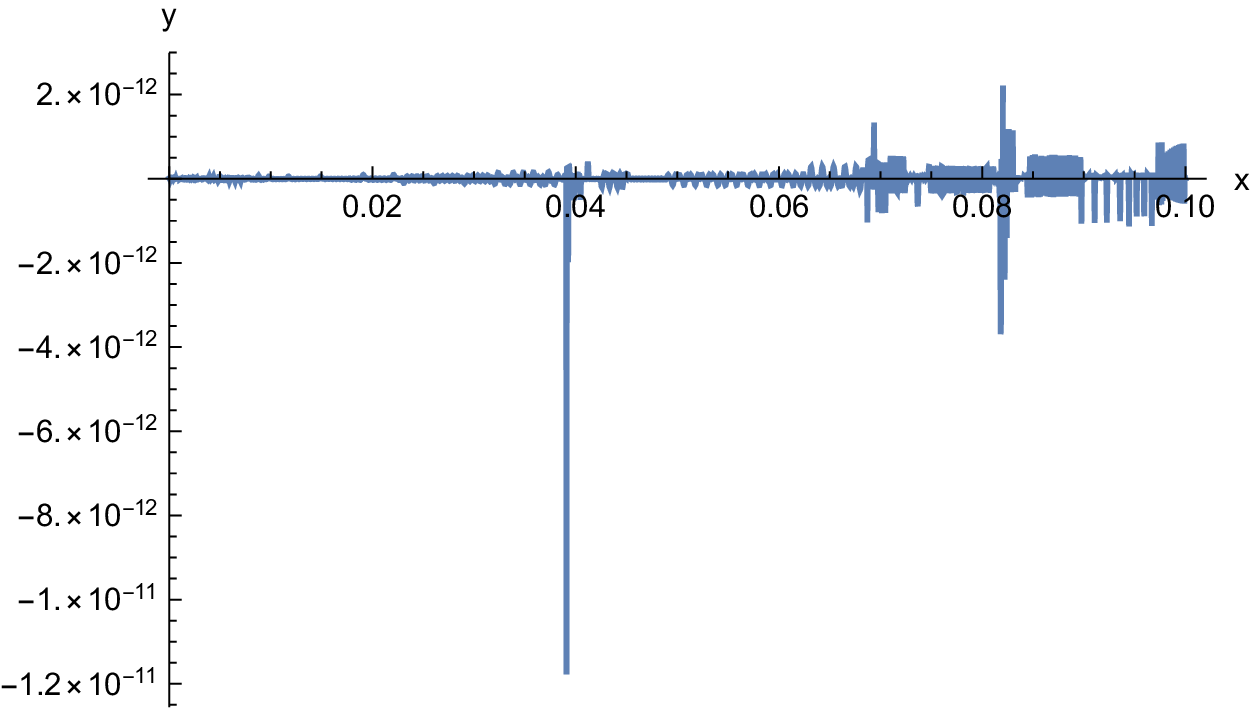}\,\,\,\,\,\,\,\,\,\,\,\,\,\,
  \includegraphics[width=2.5in]{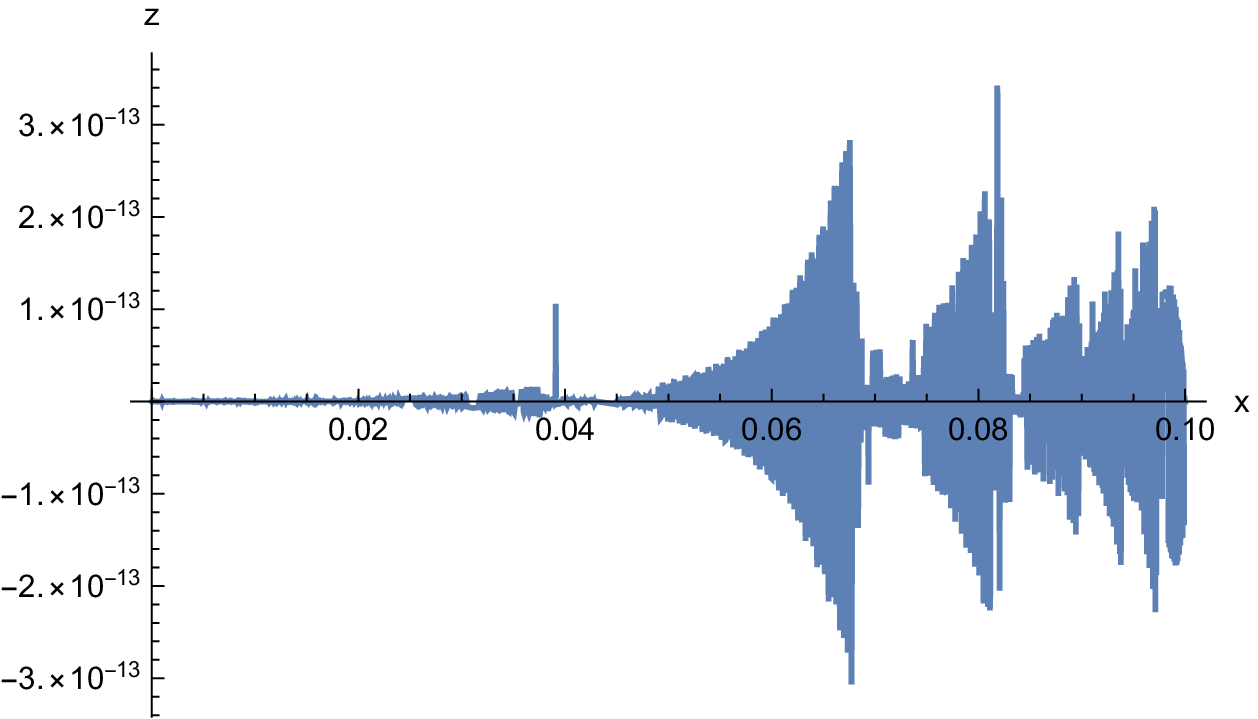}
\end{center}
 \caption{Same as in fig.~\ref{figure2}, but with $m_b=m_f=H$.}\label{figure2a}
\end{figure}

As established in \eqref{AHlate}, the apparent horizon of the vacuum state of the $\caln=2^*$ gauge 
theory in $dS_4$ occurs past the point along the 
radial holographic direction where $a_v$ vanishes. In the  radial coordinate \eqref{defx} this means that 
$x_{AH}>x_s$. So we need to extend solution of \eqref{xeoms1}-\eqref{xeoms4} past $x=x_s$. The extension is 
simple for $\{r_{2,1},c_{2,0}\}\ll 1$. Recall that to order $\calo(r_{2,1}^2,c_{2,0})$ the warp factors $a_0$ and 
$s_0$ are that of the $\caln=4$ SYM represented by \eqref{n4vsolu} --- there is nothing particular about 
$x=x_s=\frac 19$ point. Analytic continuation of \eqref{r0pert} and \eqref{c0pert} for $x>x_s$ yields:
\begin{equation}
\begin{split}
r_0=&1-r_{2,1}\
\biggl( x^2 \left((5 x-1) \arctan\frac{5 x-1}{\sqrt{(9 x-1) (1-x)}}+\sqrt{(9 x-1) (1-x)}\right)\\
&+\frac12 \pi x^2 (5 x-1)\biggr)\
 \frac{1}{(9 x-1)^{3/2} (1-x)^{3/2}}
+\calo(r_{2,1}^2,c_{2,0}^2)\,,
\end{split}
\eqlabel{r0perts}
\end{equation}    
\begin{equation}
\begin{split}
c_0=&1+c_{2,0}\ x^2 \biggl(16 \arctan\frac{5 x-1}{\sqrt{(9x-1) (1-x)}} x^2+(5x-1)\sqrt{(9x-1) (1-x)} \\
&+8 \pi x^2\biggr)^2\  \frac{1}{(9 x-1)^3 (1-x)^3}
+\calo(r_{2,1}^2,c_{2,0}^2)\,.
\end{split}
\eqlabel{c0perts}
\end{equation}    

For general values of $\{r_{2,1},c_{2,0}\}$ we need to resort to numerics. Instead of solving \eqref{xeoms1}-\eqref{xeoms2},
we solve equations \eqref{veoms} (rewritten in $x$ coordinate), using as initial condition the asymptotic behaviour 
\eqref{irass}. In this process we do not solve the constraint equations \eqref{veoms2}, but rather monitor their residuals as 
a test on our analysis. These residuals are shown in figs.\ref{figure2}-\ref{figure2a}.

\begin{figure}[t]
\begin{center}
\psfrag{x}{{$\left(x-\frac19\right)$}}
\psfrag{y}{{$r_v$}}
\psfrag{t}{{$c_v$}}
\psfrag{g}{{$\frac{m_b^2}{H^2}=0.15$}}
\psfrag{h}{{$\frac{m_b^2}{H^2}=\frac{m_f^2}{H^2}=0.15$}}
  \includegraphics[width=2.5in]{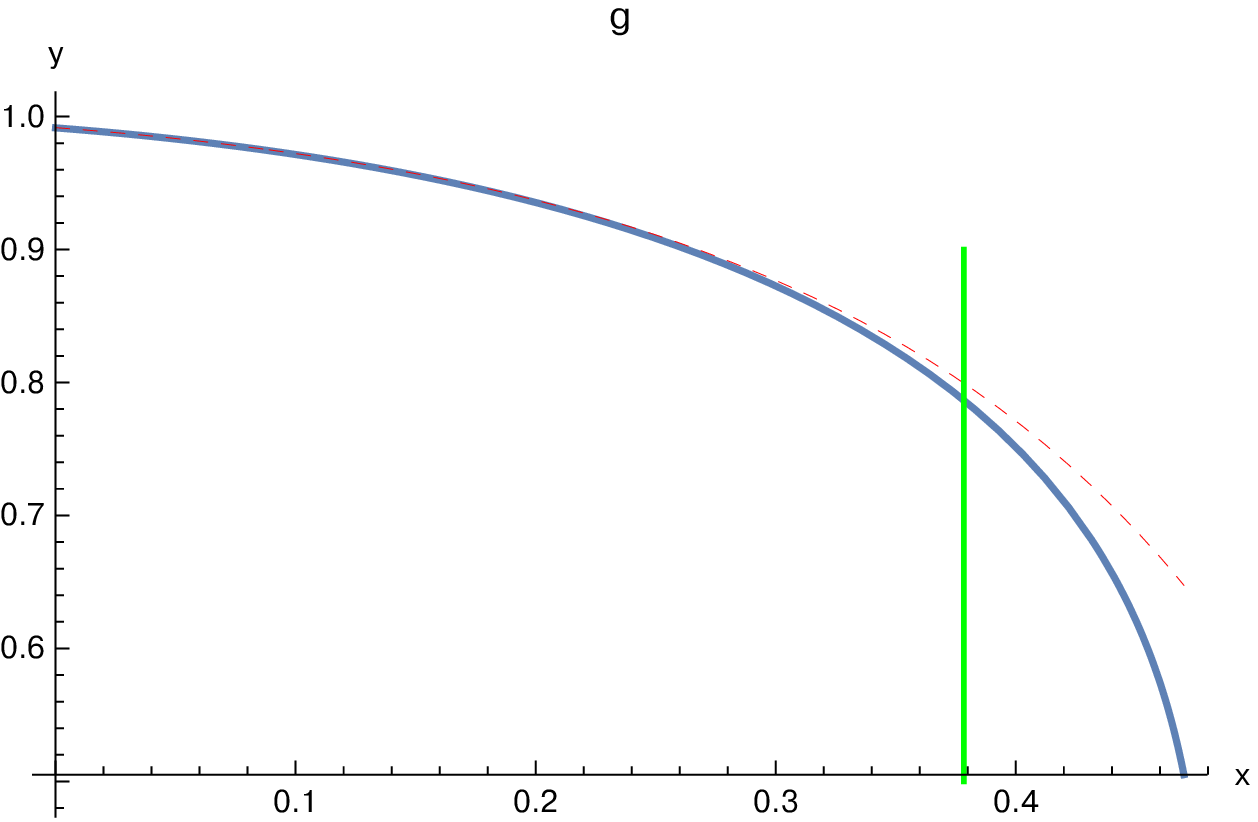}\,\,\,\,\,\,\,\,\,\,\,\,\,\,
  \includegraphics[width=2.5in]{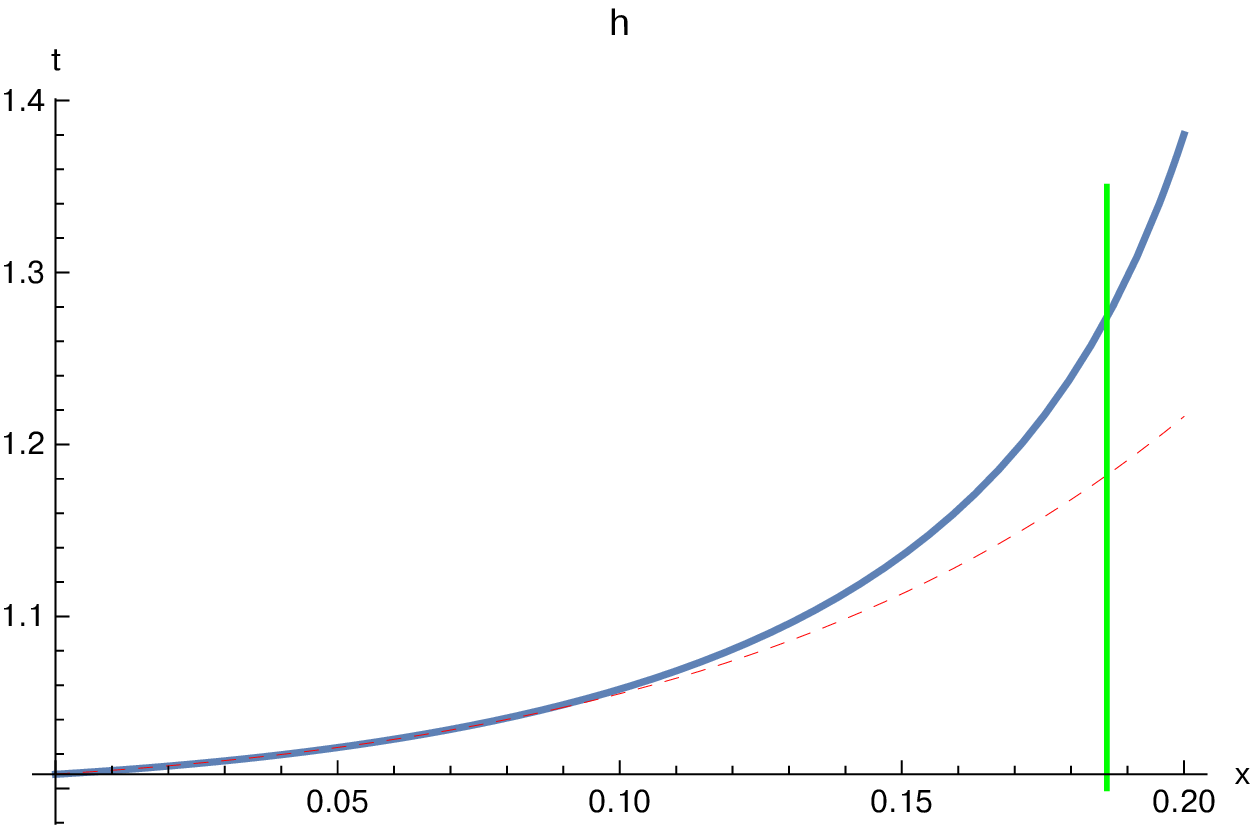}
\end{center}
 \caption{Profiles of the gravitational scalars $\{r_0,c_0\}$ (solid curves) describing the holographic dual to 
$\caln=2^*$ gauge theory vacuum in $dS_4$ for select values of $\{m_b,m_f\}$ for radial coordinate $x>x_s=\frac 19$. 
The dashed red lines represent the analytic solutions valid for $\{m_b,m_f\}\ll H$, see
\eqref{r0perts} and \eqref{c0perts}. Vertical green lines indicate the location
of the apparent horizon in the dual gravitational background.}\label{figure1}
\end{figure}

\begin{figure}[t]
\begin{center}
\psfrag{u}{{$\left(x_{AH}-\frac19\right)$}}
\psfrag{t}{{$\frac{m_b^2}{H^2}=\frac{m_f^2}{H^2}$}}
\psfrag{x}{{$\left(x-\frac19\right)$}}
\psfrag{y}{{$\frac{16x^2}{H^2}\ \lim_{t\to \infty}\frac{d_+\Sigma}{a(t)}$}}
  \includegraphics[width=2.5in]{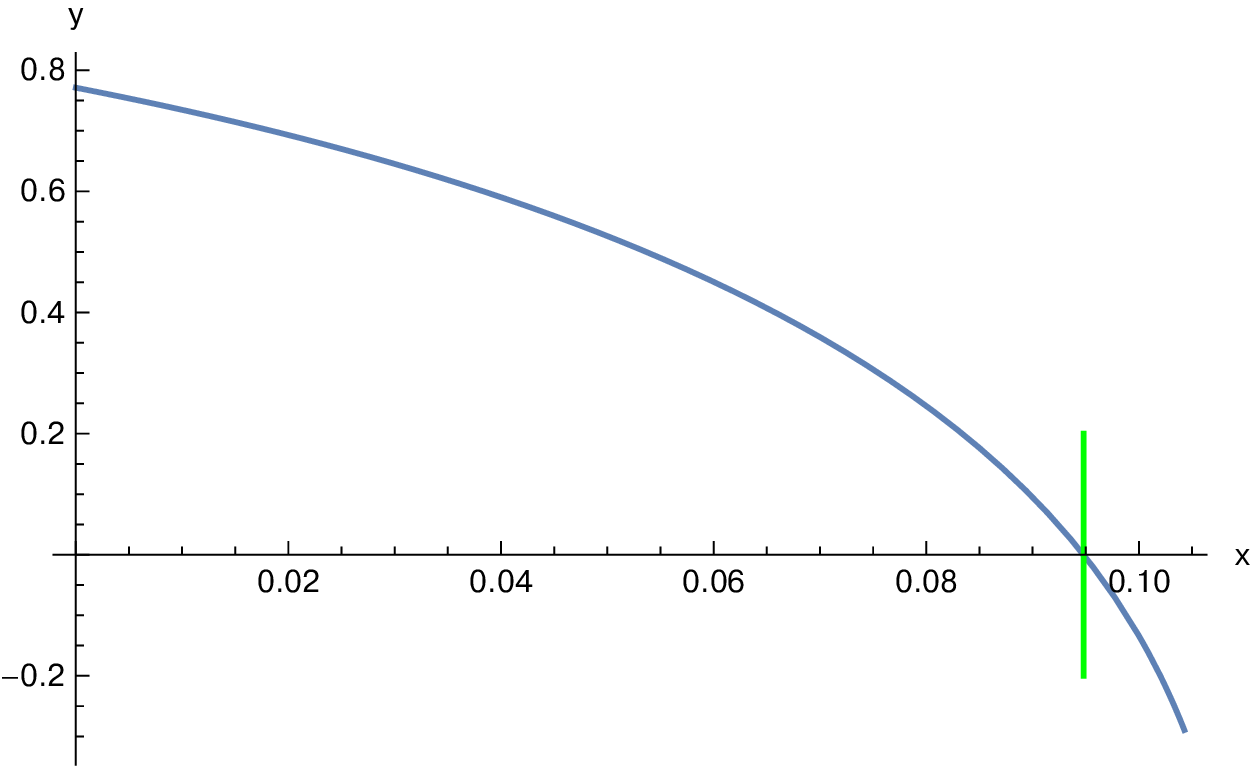}\,\,\,\,\,\,\,\,\,\,\,\,\,\,
  \includegraphics[width=2.5in]{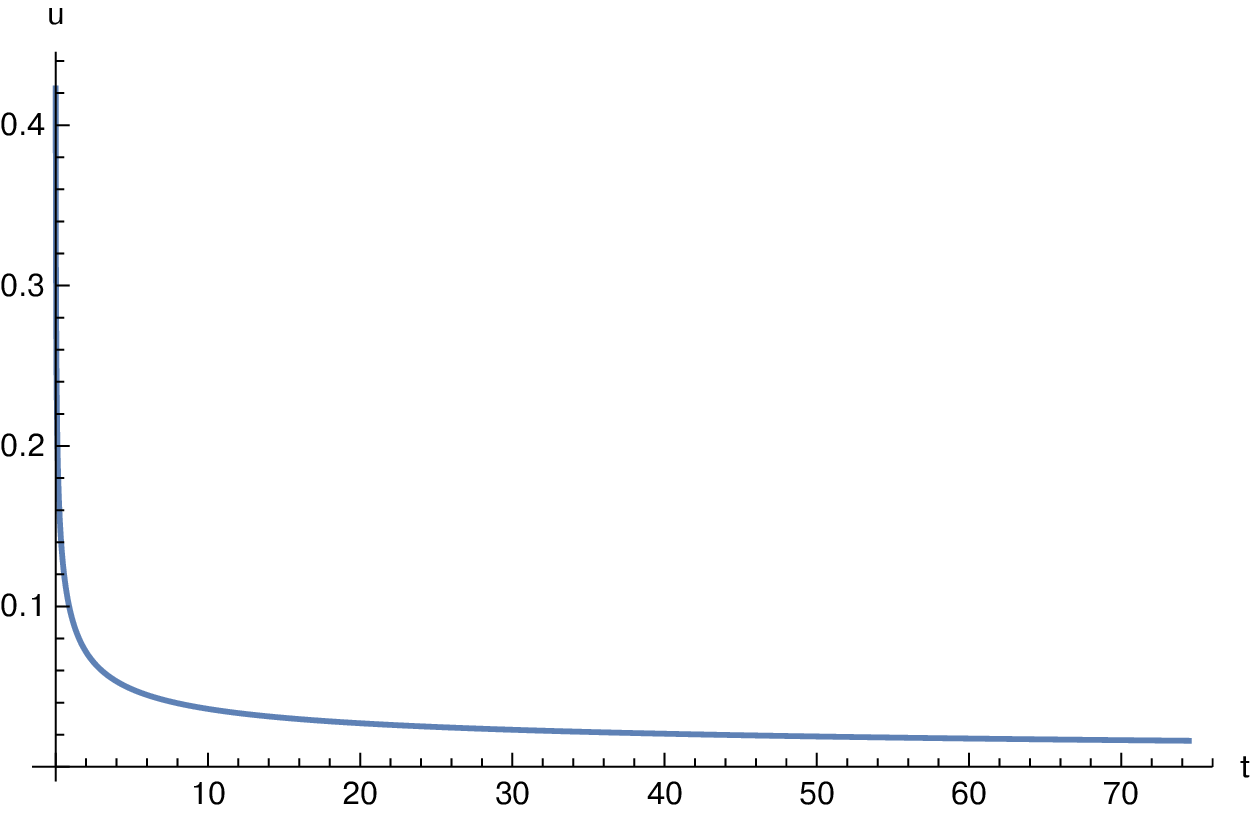}
\end{center}
 \caption{Left panel: Apparent horizon of the $\caln=2^*$ vacuum is determined from
\eqref{AHlate}. The plot represents the $d_+\Sigma$ profile of the gravitational dual
for $m_b=m_f=H$. Vertical green line indicates the location of the apparent horizon.
Right panel: location of the apparent horizon $x_{AH}$ as a function of
$m_b=m_f$. 
}\label{figure7}
\end{figure}

Fig.~\ref{figure1} represents profiles of the gravitational scalars $\{r_0,c_0\}$ for $x>x_s$,
for the same values of mass parameters as in fig.~\ref{figure1a}. Notice that the deviation
of the numerical solutions away from the perturbative results
\eqref{r0perts} and \eqref{c0perts} becomes more pronounced as one approaches
the apparent horizon, indicated by vertical green lines, \ie the gravitational
scalar fields backreaction is strong near the apparent horizon.  In general,
the apparent horizon is determined by locating the first zero in $d_+\Sigma$ for $x>x_s$,
see \eqref{AHlate}. We illustrate the procedure in the left panel of fig.~\ref{figure7}
for $\caln=2^*$ gauge theory vacuum in $dS_4$ at $m_b=m_f=H$. The right panel of
fig.~\ref{figure7} presents the location of the apparent horizon $x_{AH}$ in the gravitational
dual to $\caln=2^*$ $dS_4$ vacuum at $m_b=m_f$.

\begin{figure}[t]
\begin{center}
\psfrag{x}{{$\frac{m_b^2}{H^2}$}}
\psfrag{g}{{$m_f=0$}}
\psfrag{h}{{$m_b=m_f$}}
\psfrag{y}{{$\frac{m_b^2}{H^2}=\frac{m_f^2}{H^2}$}}
\psfrag{z}{{$\frac{16\pi}{N^2}\ \calr$}}
  \includegraphics[width=2.5in]{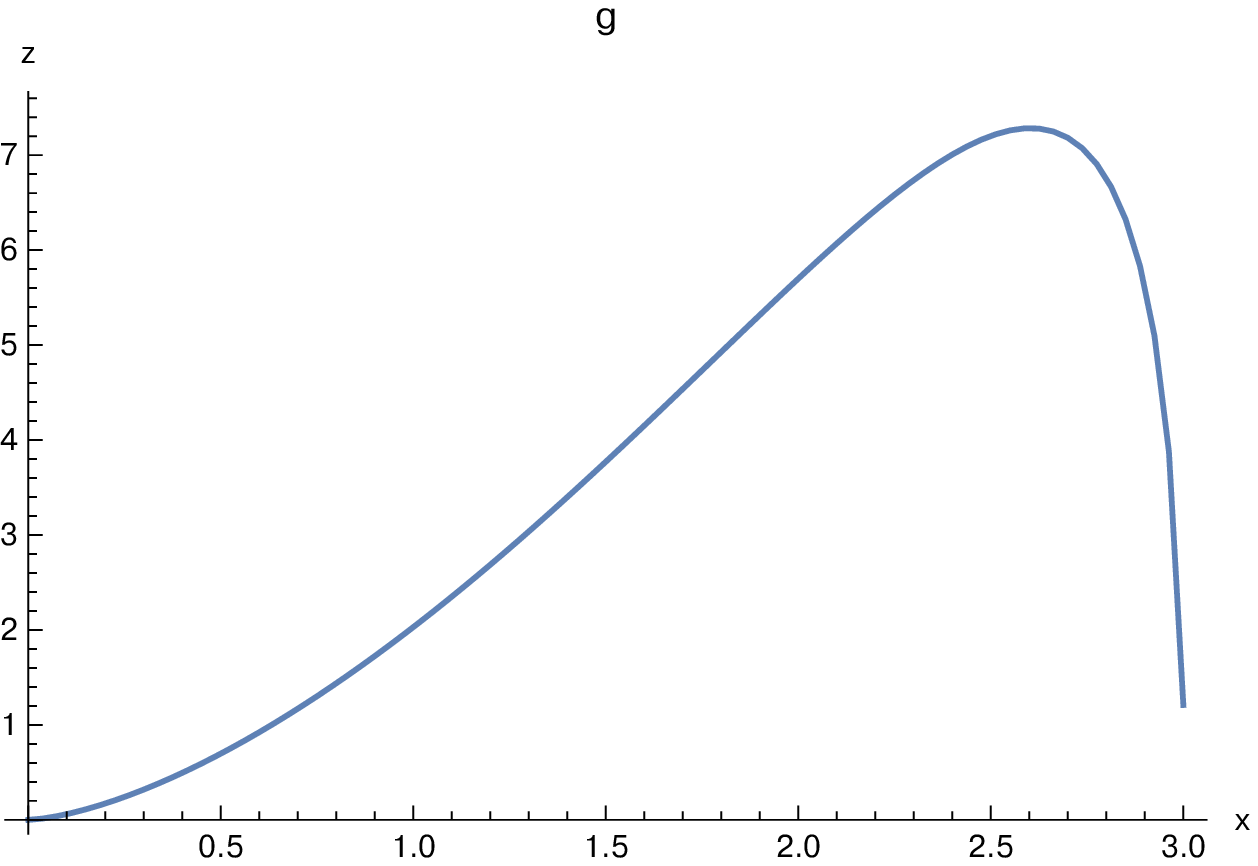}\,\,\,\,\,\,
    \includegraphics[width=2.5in]{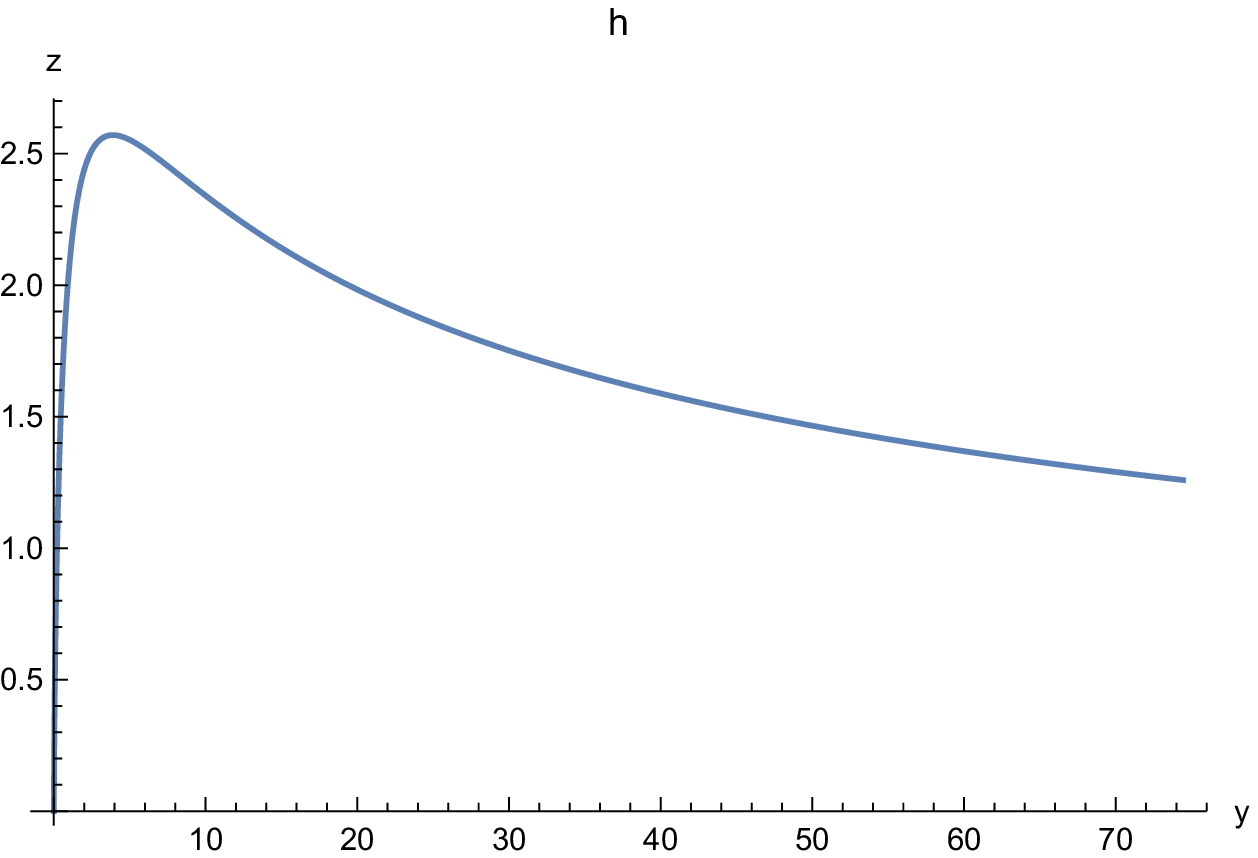}
\end{center}
 \caption{Comoving entropy production rate $\calr$ \eqref{dsdtlate} of $\caln=2^*$ gauge theory in the $dS_4$ vacuum
 for select values of $\{m_b,m_f\}$.}\label{figure5}
\end{figure}

Once we identify the location of the apparent horizon in the gravitational dual, we can compute the
comoving entropy density production rate $\calr$ of the theory in $dS_4$ vacuum, as defined in \eqref{dsdtlate},   
using \eqref{dasdt} as $t\to\infty$,
\begin{equation}
3H\times \calr=\frac{1}{H^3}\ \frac{2N^2}{\pi}\ (\sigma_v^3)'\ \frac{(A_v\chi_v')^2+(A_v\a_v')^2}{-4\calp}\bigg|_{r=r_{AH}}\,.
\eqlabel{defrhol}
\end{equation}
Alternatively, we can use \eqref{as} to deduce
\begin{equation}
\calr=\frac{1}{H^3}\  \frac{N^2}{16\pi}\ (\sigma_v)^3\bigg|_{r=r_{AH}}\,.
\eqlabel{altr}
\end{equation}
Both formulas should agree --- in fact, we find that the relative difference
between the two results over the range of the mass parameters studied
is $\lesssim 10^{-10}$. This excellent agreement is  expected given the consistency checks
on the numerics presented in figs.~\ref{figure2}-\ref{figure2a}.
Results for the comoving entropy production rate $\calr$ are presented in fig.~\ref{figure5}.

\section{Conclusion}\label{conclude}

While a full understanding of quantum gravity is lacking, inflationary models can
be improved with  {\it quantum} treatment of the Standard Model
gauge theories in {\it classical} gravitational backgrounds.
This approximation is justified when there is a large hierarchy between the
relevant QFT scales and the Planck scale. Thus, one needs to understand a QFT stress-energy
tensor in curved space-time, see \eqref{einstein}. This subject has a long history: although it is
a textbook material \cite{Birrell:1982ix}, there are many unsolved problems, even in the case of
weakly-coupled theories.

In this paper we build up on the early observation 
\cite{Buchel:2002wf} that when applicable, the holographic gauge/gravity correspondence
\cite{m1,Aharony:1999ti} can be used to achieve this goal. The Standard Model
gauge theories do not have weakly coupled holographic gravitational dual; however,
explicit holographic models can be studied to deduce the features of strongly coupled
gauge theories in curved space-time. Here, we used the $\caln=2^*$ holography
to, hopefully, extract the general properties of quantum gauge theories in
$dS_4$ space-time.

We summarize now our main observations.
\nxt We argued that an initial state of the gauge theory in $dS_4$ would evolve to
a universal static state which is a Bunch-Davies (Euclidean) vacuum. It does not appear possible
to holographically evolve to a MA-vacuum \cite{Mottola:1984ar,Allen:1985ux}.
This supports the claims in the literature
\cite{Banks:2002nv,Einhorn:2002nu,Einhorn:2003xb,Collins:2003zv} that interactions in QFTs
invalidate the free-field construction of MA-vacua.
\nxt We presented a holographic analog of the Weyl transformation between the Minkowski
and the FLRW space-times. A conformal theory, up to computable anomalies, is invariant
under this Weyl transformation; for a non-conformal theory one can map
FLRW dynamics to Minkowski dynamics, provided the coupling constants of the relevant operators
in the theory are appropriately ``quenched'' --- prescribed a particular time-dependence.  
\nxt Holography provides a natural concept of entropy of a system far from equilibrium.
Specifically, one identifies the entropy of the boundary gauge theory with the
Bekenstein-Hawking entropy of the apparent horizon in the gravitational dual
\cite{Booth:2005qc,Figueras:2009iu}. Apparent horizon is observer dependent, however,
in holographic setting describing dynamical evolution of spatially  homogeneous and
isotropy states of the boundary theory, there is a ``preferred slicing'' of the dual geometry.
When the dynamical evolution of the gauge theory is hydrodynamic, the
non-equilibrium entropy, defined in this manner, correctly reproduces Landau entropy
\cite{ll}. We proved the theorem that even far from equilibrium, entropy associated
with the apparent horizon of the gravitational holographic dual does not decrease with time.
We also showed that the comoving entropy production rate computed with respect to a
conformal time is invariant under Weyl transformations --- there is no 'conformal anomaly'
for the entropy transformations.
\nxt Perhaps most surprisingly, we discovered that while the Bunch-Davies vacuum of a
CFT is adiabatic --- there is no comoving entropy production, a non-conformal gauge theory
has a constant physical entropy density in its vacuum state; as a result the comoving
entropy density grows with the background geometry scale factor as $a(t)^3$.  
\nxt One of the motivations of the project was an explicit computation of the gauge theory
vacuum energy density $\cale_v$, to address the feasibility of self-consistent and reliable
accelerated expansion of the
Universe, driven by the latter. Here, 'reliable' means that the solution of the Friedmann
equation \eqref{desitter2} produces the Hubble constant $H\ll M_{pl}$. Results for the vacuum
energy density computation for $\caln=2^*$ gauge theory at strong coupling are presented in
figs.~\ref{figure3}-\ref{figure4}. Much as it is  in the case for the computation of the stress-energy
tensor of weakly coupled QFTs in curved space-time, $\cale_v$ is prone to renormalization
scheme ambiguities. The necessary condition for the accelerated expansion,
\ie $\cale_v>0$ can be claimed in a scheme-independent way only for small values of the
mass parameters $\{m_b,m_f\}\lesssim H$. However, in this case we are close to a conformal fixed point
($\caln=4$ SYM in this case), and Friedmann equation produces $H\sim M_{pl}$. To conclude,
$\caln=2^*$ gauge theory at strong coupling can not produce accelerated expansion
of the Universe with classical gravity.

Can the 'reliability' aspect of the accelerated expansion of the Universe from the
$\caln=2^*$ vacuum energy discussed be fixed? Let's speculate that the vacuum energy of a QFT 
is (motivated by \eqref{caleres})
\begin{equation}
 \cale_v(H)=Q_1^2 H^4+ \biggl(Q_2 (m_b^4-m_f^4)-Q_3^2 m_f^2 H^2\biggr)\ln\frac{H}{\Lambda}
+q_1 m_f^2 H^2+q_2 m_f^4+q_3 m_b^2 H^2+q_4 m_b^4\,,
\eqlabel{fantacy1}
\end{equation}
where $Q_i$ are fixed constants, and $\Lambda$ and $q_i$ encode scheme dependence.
Basically, \eqref{fantacy1} maintains the rough features of the supersymmetry of the
$\caln=2^*$ model (whether $m_b=m_f$ or $m_b\ne m_f$), the ultraviolet conformal
fixed point contribution ( $Q^2_1 H^4$ term ), and the structure of the scheme dependence.
What is different from the $\caln=2^*$ model is the absence of contributions to
the vacuum energy from the condensates of the relevant operators (the $r_{2,0}$
and $c_{4,0}$ terms in \eqref{caleres}).  Let's analyze \eqref{fantacy1} in the context
of the Friedmann equation \eqref{desitter2}. We are looking for a solution to \eqref{desitter2}
with $H\ll M_{pl}$. The renormalization scale $\Lambda$ is a QFT feature, so it is natural to have
$\Lambda\ll M_{pl}$. There is no canonical way to define the energy density
of a non-supersymmetric QFT ---
this ambiguity is reflected in $q_2$ and $q_4$ coefficients; however, if the theory is
supersymmetric in Minkowski space-time, one expects that
\begin{equation}
\lim_{H\to 0}\ \cale_{v}(H)\bigg|_{m_b=m_f=m} = 0\,,
\eqlabel{fantacy2}
\end{equation}
provided the $H\to 0$ limit is smooth. Note that this limit is indeed smooth with
$m_b=m_f$. Restricting to supersymmetric theories in Minkowski, we thus require $q_2+q_4=0$.
Thus,
\begin{equation}
 \cale_v^{susy}(H)=Q_1^2 H^4+ \biggl(-Q_3^2 m^2 H^2\biggr)\ln\frac{H}{\Lambda}
+(q_1+q_3)\ m^2 H^2 \,.
\eqlabel{fantacy3}
\end{equation}
It is not possible to further fix the ambiguity associated with $(q_1+q_3)$ ---
even for a Euclidean supersymmetric formulation of  gauge theories on $S^4$
such a term can not be fixed \cite{Bobev:2013cja}.
Interestingly, if $H\ll \Lambda$, the renormalization scheme sensitivity can be made arbitrarily
weak. Moreover, if $H\ll m$, the conformal fixed point contribution is also irrelevant.
The latter statement is precise if we choose the renormalization scale of the theory $\Lambda=m$.
Thus,
\begin{equation}
 \cale_v^{susy}(H)\approx -Q_3^2\ m^2 H^2 \ln\frac{H}{\Lambda}\,,\qquad  H\ll m\ll M_{pl}\,.
\eqlabel{fantacy4}
\end{equation}
Solving the Friedmann equation \eqref{desitter2} with the vacuum energy density
\eqref{fantacy3} we find
\begin{equation}
H= m\ \exp\biggl[-\frac{3M_{pl}^2}{8\pi Q_3^2 m^2}\biggr]\,,
\eqlabel{reshub}
\end{equation}
naturally enforcing the hierarchy of scales in  \eqref{fantacy3}.
As we emphasized, unfortunately, the $\caln=2^*$ holographic model is {\it not}
within the class of models with the energy density given by \eqref{fantacy1} --- in this theory
in the limit $m\gg H$ the contribution of the condensates of the relevant operators become important.
In fact, precisely for $m_b=m_f$ case it can be shown that for $\frac{m}{H}\to \infty$
the effective coefficient $Q_3$ in $\caln=2^*$ model vanishes.  

It is interesting to further develop the dynamics of gauge theories in
de Sitter. What replaces the 'hydrodynamics' in the approach of a QFT
to its Bunch-Davies vacuum? What is the microscopic origin of the de Sitter entropy
density of non-conformal theories? Do holographic models in the class
\eqref{fantacy1} exist? We hope to address these questions in the future.

\section*{Acknowledgments}
Research at Perimeter
Institute is supported by the Government of Canada through Industry
Canada and by the Province of Ontario through the Ministry of
Research \& Innovation. This work was further supported by
NSERC through the Discovery Grants program (AB).

\appendix

\section{Apparent horizon area growth theorem}\label{th1}

We prove here that the area density of the apparent horizon $\cala_{AH}$, see \eqref{as},
\begin{equation}
\cala_{AH}(t)\equiv \Sigma(t,r)^3\bigg|_{r=r_{AH}}\,,
\eqlabel{pr1}
\end{equation}
in a gravitational dual to a dynamical evolution of the boundary $\caln=2^*$ gauge theory state does not decrease 
with the boundary time $t$,
\begin{equation}
\frac{d\cala_{AH}}{dt}\ge 0\,.
\eqlabel{pr2}
\end{equation}
This theorem  implies that the comoving entropy density of the boundary gauge theory state, identified with the 
the Bekenstein-Hawking entropy density of the apparent horizon would satisfy the second law of 
thermodynamics. 

Recall that \eqref{dasdt}
\begin{equation}
\frac{d\cala_{AH}}{dt}=8\ (\Sigma^3)'\ \frac{
 (d_+\chi)^2+3 (d_+\a)^2)}{- \calp}\bigg|_{r=r_{AH}}.
\eqlabel{dasdt3}
\end{equation}
Thus we need to establish two facts:
\begin{itemize}
\item $\Sigma'(t,r_{{AH}}+0)\ge 0$;
\item $\calp\bigg|_{r={r_{AH}}+0}\le 0$.
\end{itemize}

Integrating the constraint equation \eqref{ham} and using the boundary condition \eqref{bcdata} we conclude
\begin{equation}
\Sigma'(t,r)=\frac a2+\int_r^\infty\ d\r\  \frac 43\ \Sigma(t,\r)\ \biggl( \left(\frac{\del\chi(t,\r)}{\del\r}\right)^2
+3 \left(\frac{\del\a(t,\r)}{\del\r}\right)^2\biggr)\,.
\eqlabel{pr3}
\end{equation} 
Since $\Sigma(t,r)>0$ during the evolution, $\Sigma'(t,r)\ge 0$. 

Apparent horizon is defined as the innermost (with respect to the boundary) radial coordinate location $r=r_{AH}$, where $d_+\Sigma(t,r)$ 
vanishes. Note from \eqref{bcdata} that 
\begin{equation}
d_+\Sigma(t,r) = \frac{a r^2}{16}+\calo(r^0)\ >0\,,\qquad r\to\infty \,.
\eqlabel{dpinfty}
\end{equation}
Thus, assuming analyticity of the background geometry,
\begin{equation}
d_+\Sigma(t,r)> 0\,,\qquad r> r_{AH}\qquad \Longrightarrow\qquad d_+'\Sigma(t,r)\bigg|_{r=r_{AH}}\ge 0\,.
\eqlabel{pr4}
\end{equation}
The first evolution equation in \eqref{ev1} evaluated at the apparent horizon reads 
\begin{equation}
0=d'_+\Sigma+\frac{\Sigma}{6}\ \calp\bigg|_{r=r_{AH}+0}\qquad \Longrightarrow\qquad \calp_{r=r_{AH}+0}\le 0\,.
\eqlabel{pr5}
\end{equation}

\section{Mapping QFT dynamics in Minkowski and FLRW space-times}\label{th2}

We argue that holographic QFT dynamics of homogeneous and isotropic states in Friedmann-Lemaitre-Robertson-Walker
(FLRW) Universe 
\begin{equation}
ds^2=-dt^2 + a(t)^2\ d\boldsymbol{x}^2\,,
\eqlabel{flrwm}
\end{equation}
is equivalent to a quenched dynamics \cite{Buchel:2012gw} of the same theories in Minkowski space-time 
\begin{equation}
d\hat{s}^2=-d\tau^2 + d\boldsymbol{x}^2\,,\qquad \t=\int_0^t \frac{d\nu}{a(\nu)}\,,
\eqlabel{flrw2}
\end{equation}
where any relevant\footnote{It is possible to extend the result to marginal couplings as well.} 
coupling constant $\l_\Delta$ of dimension $\Delta < 4$ operator $\calo_\Delta$ being replaced with the 
corresponding time-dependent coupling $\hat{\l}_\Delta$ according to 
\begin{equation}
\l_\Delta\ \to\ \hat{\l}_\Delta(\t) = a(t(\t))^{4-\Delta} \l_\Delta\,.
\eqlabel{transforml}
\end{equation}

The proof is elementary --- while we implicitly focus on $\caln=2^*$ model of holography,
it  is easily generalized to other holographic dualities, including those with gravitational bulk gauge fields 
(corresponding to charged states of the boundary gauge theory). The only requirement is that the QFT 
has an ultraviolet conformal fixed point. Indeed, the gravitational metric dual to QFT dynamics 
of homogeneous and isotropic states is given by \eqref{EFmetric}, with the bulk scalars $\phi_\Delta=\phi_\Delta(t,r)$,
dual to operators $\calo_\Delta$. The asymptotic boundary expansion encodes the background gauge theory 
metric and the coupling constants $\l_\Delta$ (compare with \eqref{bcdata}):
\begin{equation}
\Sigma=\frac{a(t) r }{2}+\calo(r^{-1})\,,\qquad A=\frac{r^2}{8}-\frac{\dot{a}(t) r}{a(t)}+\calo(r^0)\,,\qquad 
\phi_\Delta=\frac{\l_\Delta}{r^{4-\Delta}}+\calo(r^{5-\Delta})\,,
\eqlabel{bcflrw}
\end{equation}
where we used the fact that QFT is a four-dimensional CFT in the ultraviolet. 
Holographic equations of motion --- the analog of \eqref{ev1}-\eqref{mom} --- come from the 
diffeomorphism  invariant effective action in five dimensions. Thus, they are covariant under
arbitrary change of bulk coordinates 
\begin{equation}
\{t,r\}\ \to \{\tau(t,r),\rho(t,r)\}\,.
\eqlabel{change}
\end{equation}
Let's choose 
\begin{equation}
\t\equiv \int_0^t \frac{d\nu}{a(\nu)}\,,\qquad \r\equiv r a(t)\,,
\eqlabel{deftrho}
\end{equation}
and introduce $\{\hat{\Sigma},\hat{A},\hat{\phi}_\Delta\}$ as 
\begin{equation}
\begin{split}
&\hat{\Sigma}(\t,\r)\equiv\Sigma(t(\t),\r/a(t(\t)))\,,\\
&\hat{A}(\t,\r)\equiv a(t(\t))^2 A(t(\t),\r/a(t(\t)))+\frac{\r}{a(t(\t))}\ \frac{d}{d\t}a(t(\t))\,,\\
&\hat{\phi}_\Delta(\t,\r)\equiv \phi_\Delta(t(\t),\r/a(t(\t)))\,.
\end{split}
\eqlabel{warptr}
\end{equation}
We emphasize that \eqref{deftrho} and \eqref{warptr} is nothing but the change of coordinates, 
accompanied by field redefinitions. While mathematically trivial, it has profound implications 
physically:   
transformation \eqref{warptr} reparameterizes the background metric \eqref{EFmetric} $ds_5^2$ as 
\begin{equation}
ds_5^2\ \to d\hat{s}_5^2\equiv ds_5^2= 2 d\t\ (d\r-\hat{A} d\t)+\hat{\Sigma}^2\ d\boldsymbol{x}^2\,,
\eqlabel{EFnew}
\end{equation}
and produces a new set of the boundary conditions 
\begin{equation}
\hat{\Sigma}=\frac{\r }{2}+\calo(\r^{-1})\,,\qquad \hat{A}=\frac{\r^2}{8}+\calo(\r^0)\,,\qquad 
\hat{\phi}_\Delta=\frac{\hat{\l}_\Delta(\t)}{\r^{4-\Delta}}+\calo(\r^{5-\Delta})\,,
\eqlabel{bcflrw2}
\end{equation}
where $\hat{\l}_\Delta(\t)$ is given by \eqref{transforml}. \eqref{EFnew} and \eqref{bcflrw2} imply that the 
transformed holographic dynamics is that of the boundary gauge theory in Minkowski space-time 
\eqref{flrw2} with quenched relevant couplings given by \eqref{transforml}. Note that 
\eqref{warptr} applied at $\t=t=0$ maps the initial states of the theories for equivalent dynamics 
in  FLRW and $R^{3,1}$.  

Interestingly, the map \eqref{warptr} leaves invariant the rate of the comoving entropy production 
computed with respect to the conformal time $d\t=\frac{dt}{a(t)}$. Indeed, 
since 
\begin{equation}
\begin{split}
&d_+\equiv \del_t+A(t,r)\ \del_r= \frac{1}{a(t(\t))}\left(\del_\t +\hat{A}(\t,\r)\ \del_\r\right) = \frac{1}{a(t(\t))}\ \hat{d}_+\,,\\
&\del_r= a(t(\t))\ \del_\r\,,
\end{split}
\eqlabel{dptrans}
\end{equation}
\eqref{dasdt2} implies that
\begin{equation}
\frac{d(a^3s)}{dt} = \frac{1}{a(t(\t))}\ \frac{d \hat{s}}{d\t}\,.
\eqlabel{entrate2}
\end{equation}
Notice that \eqref{entrate2} establishes that equilibrium states of CFTs in Minkowski space-time are adiabatic in de Sitter.

\section{$\caln=4$ SYM in FLRW}\label{n4ds}

We consider here a restrictive set of  homogeneous and isotropic states of $\caln=4$ SYM in FLRW space-time --- the ``thermal'' 
states \cite{Apostolopoulos:2008ru,Buchel:2016cbj}. 
These states are represented by solutions of holographic equations \eqref{ev1}-\eqref{mom} with gravitational 
scalars $\a$ and $\chi$ identically set to zero.  

The general solution of the holographic equations of motion within $\a=\chi\equiv 0$ ansatz takes form 
\begin{equation}
\begin{split}
&A=\frac{(r+\l(t))^2}{8}-(r+\l(t))\ \frac{{\dot a}(t)}{a(t)}-{\dot\lambda}(t)-\frac{\mu^4}{8a(t)^4(r+\l(t))^2}\,,\\
&\Sigma=\frac{(r+\l(t))a(t)}{2}\,,
\end{split}
\eqlabel{n4solve}
\end{equation}
where $\l(t)$ is an arbitrary function related to the residual diffeomorphism invariance of the metric \eqref{EFmetric},
and $\mu$ is a constant. Without loss of generality we assume that $\l<0$. From \eqref{defhorloc}, 
the apparent horizon is then located at 
\begin{equation}
r_{AH}=-\l(t)+\frac{\mu}{a(t)}\,.
\eqlabel{n4ah}
\end{equation}
The comoving entropy density of the gauge theory is (see \eqref{as})
\begin{equation}
a^3 s= \frac{N^2}{16\pi}\ \frac{\mu^3}{8}\,.
\eqlabel{asn4}
\end{equation}
Note that in agreement with \eqref{dasdt}, the comoving entropy density does not 
change with time; in particular, the entropy production rate $\calr$  \eqref{dsdtlate} in the vacuum state  vanishes.

The adiabaticity of the evolution \eqref{n4solve} can be understood from the equivalence theorem of Appendix \ref{th2}.   
Choosing the gauge $\l(t)=0$ (as used in the derivation of the theorem) and applying the map \eqref{warptr}, we find 
\begin{equation}
\hat{\Sigma}=\frac \r2\,,\qquad \hat {A}=\frac{\r^2}{8}\left(1-\frac{\mu^4}{\r^4}\right)\,,
\eqlabel{n4mink} 
\end{equation}
which is the $AdS_5$ Schwarzschild black brane metric, dual to the equilibrium 
thermal state of the $\caln=4$ SYM. There is no entropy production in the thermal state;
correspondingly, following the theorem of   Appendix \ref{th2}, there is no comoving entropy 
production in the state \eqref{n4solve}. It is for this reason we refer to the latter state as "thermal''. 

We would like to stress, as the $\caln=4$ SYM example illustrates here, that even though the  equivalence theorem of Appendix \ref{th2}
maps the dynamics of a QFT in FLRW to the one in Minkowski space-time, the energy density of the theory in Minkowski space-time 
is not the comoving energy density of the theory in FLRW: 
\begin{equation}
\begin{split}
\hat\cale=\frac38 \pi^2 N^2 \left(\frac{\mu}{4\pi}\right)^4\,,\qquad 
a(t)^4\ \cale(t)=\hat\cale + \frac{3N^2({\dot a}(t))^4}{32\pi^2}\,.
\end{split}       
\eqlabel{energydensity}
\end{equation}
The difference between the two is due to the conformal anomaly in transformation from Minkowski to FLRW 
space-time \cite{Apostolopoulos:2008ru}.
 
Thermal states of $\caln=4$ SYM in $dS_4$ at late times approach Bunch-Davies vacuum. 
Choosing the gauge $\lambda(t)=-H$ and redefining a radial coordinate as $r\equiv \frac{H}{x}$, $x\in (0,x_{AH}]$
corresponding to  $r\in [r_{AH},\infty)$,
at late times $t\to \infty$, we find
\begin{equation}
\begin{split}
&{x_{AH}}\ \to\ 1\,,\\
&\frac{\Sigma}{a} \to\ \sigma_v=\frac{H}{2x}\ (1-x)\,, \\
&A\ \to\ A_v=\frac{H^2}{8 x^2}\ (1-x)(1-9x)\,.
\end{split}
\eqlabel{n4vsolu}
\end{equation}
Notice\footnote{See discussion following \eqref{AHlate}.} that $A_v$ vanishes at 
\begin{equation}
x_s=\frac 19\in (0,x_{AH}]\,,\qquad A_v(x_s)=0\,.
\eqlabel{defxs}
\end{equation} 
The $\caln=4$ SYM $dS_4$ vacuum energy is independent of the initial energy density,
and is completely determined by the conformal anomaly:
\begin{equation}
\frac{32\pi^2}{N^2}\ \frac{\cale_v}{H^4} = 3\,.
\eqlabel{n4ev}
\end{equation}

\end{document}